\documentstyle[multicol,aps,epsfig,amsbsy]{revtex}

\newcommand{\lcm}{{\rm lcm}}

\newtheorem{definition}{Definition}
\newtheorem{theorem}{Theorem}

\newtheorem{lemma}{Lemma}
\newtheorem{corollary}{Corollary}
\newtheorem{conjecture}{Conjecture}

\begin{document}

\title{Upper Bound on the Products of\\
Particle Interactions in Cellular Automata}
\author{Wim Hordijk,
Cosma Rohilla Shalizi,\thanks{Permanent address: Physics Department,
University of Wisconsin, Madison, WI 53706.} and James P. Crutchfield}
\address{Santa Fe Institute, 1399 Hyde Park Road, Santa Fe, NM 87501\\
Electronic address: \{wim,shalizi,chaos\}@santafe.edu}
\date{30 January 2001}
\maketitle

\begin{abstract}
Particle-like objects are observed to propagate and interact in many spatially
extended dynamical systems. For one of the simplest classes of such systems,
one-dimensional cellular automata, we establish a rigorous upper bound on the
number of distinct products that these interactions can generate.  The upper
bound is controlled by the structural complexity of the interacting
particles---a quantity which is defined here and which measures the amount of
spatio-temporal information that a particle stores. Along the way we establish
a number of properties of domains and particles that follow from the
computational mechanics analysis of cellular automata; thereby elucidating why
that approach is of general utility. The upper bound is tested against several
relatively complex domain-particle cellular automata and found to be tight.



\end{abstract}

\begin{multicols}{2}

\tableofcontents

\section{Introduction}

Persistent, localized, propagating structures---particles---have long been
observed and constructed in cellular automata (CAs)
\cite{Burks-essays,Winning-Ways,Peyrard-Kruskal,Grassberger-diffusion,Boccara-Nasser-Roger-particle-like,Boccara-Roger-block-transformations,Boccara-transformations,Aizawa-Nishikawa-Kaneko-soliton-turbulence,Park-Steiglitz-Thurston-soliton,Wolfram-theory-and-applications,Wolfram-CA-and-complexity,Lindgren-Nordahl-universal-computation,JPC-MM-PNAS,Yunes-firing-squad,Eloranta-partially-permutive,Eloranta-defect-ensembles,Eloranta-Nummelin-random-walk,Cellular-Automata-and-Modeling,Andre-Bennett-Koza,Attractor-basin-portrait,Hanson-thesis,Comp-mech-of-CA-example}.
A review of the literature suggests that particles are widely felt to be some
of the more interesting phenomena displayed by those systems
\cite{Eppstein-glider-rule-database}. They are analogous to the ``defects'' or
``coherent structures'' of pattern formation processes in condensed matter
physics
\cite{Grassberger-diffusion,Boccara-Nasser-Roger-particle-like,Aizawa-Nishikawa-Kaneko-soliton-turbulence,Park-Steiglitz-Thurston-soliton,Eloranta-defect-ensembles,Cellular-Automata-and-Modeling}.
In fact, in cellular automata used to model pattern formation processes, the
particles model defects and vice versa
\cite{Manneville-dissipative-structures,Cross-Hohenberg,Winfree-geometry,Winfree-time-breaks-down}. A
different analogy to condensed matter physics (specifically, hydrodynamics
\cite{Infeld-Rowlands}) gives them the name ``solitons''
\cite{Aizawa-Nishikawa-Kaneko-soliton-turbulence,Park-Steiglitz-Thurston-soliton}.
They are also known as {\em gliders}, {\em glider-like objects}, or {\em
spaceships} particularly, but not exclusively, in the context of the Game of
Life two-dimensional CA \cite{Winning-Ways,Poundstone-recursive}.  The name
``particle'', while inspired by an analogy to field theory in physics, is used
here merely for the sake of uniform terminology and neutrality of
associations.\footnote{ To avoid confusion, we should say that a ``particle''
in our sense is \textit{not} the same as a particle in the sense of interacting
particle systems (IPSs)
\cite{Griffeath-particle-systems,Liggett-particle-systems} or lattice gases
\cite{Rothman-Zaleski-text}. The particles of an IPS or the coherent structures
that emerge in lattice gases \textit{may} be particles in our sense; we hope to
explore these and related issues elsewhere.}

CA particles, like their physical counterparts, interact via collision, and
these interactions are well known to play a crucial role in the dynamics of
their underlying cellular automaton. The construction of computational devices
in CAs, for instance, is almost always accomplished through engineering the
proper interactions among particles
\cite{Burks-essays,Lindgren-Nordahl-universal-computation,JPC-MM-PNAS,Steiglitz-Kamal-Watson,Griffeath-Moore-LwoD,Moore-majority-vote,Moore-Nordahl-lattice-gases,Das-MM-JPC-discovery-of-particles,Wim-MM-JPC-mechanisms,Margolus-crystalline,Das-thesis,Wim-thesis}. Indeed,
it was even at one time conjectured by Wolfram
\cite{Wolfram-universality-and-complexity} that the presence of particles in a
CA was tantamount to its being computation-universal.  It is therefore of
considerable interest to know what interactions a CA's particles may
have. Acquiring that knowledge is significantly simplified if we can place a
bound on the number of different interactions between any pair of
particles. The first successful attempt to do so was an expression given in
\cite{Park-Steiglitz-Thurston-soliton} for particles interacting on a
completely uniform, quiescent background. It has been appreciated for some
time, however, that many CAs display patterned or textured
backgrounds---sometimes called ``domains''. Here we substantially generalize
the original formula to accommodate a large class of domains and prove the
generalization using elementary automata and number theory.

The outline of the paper is as follows.  First, we fix basic notation relating
to cellular automata, regular languages, and finite-state transducers. We then
define domains, in particular periodic domains, and prove some basic results
about them.  We next define particles as a particular kind of interface between
domains, and define interactions between particles.  After establishing some
auxiliary number-theoretic results, we prove the upper bound formula that
generalizes the main theorem of \cite{Park-Steiglitz-Thurston-soliton} to
arbitrary periodic domains. Result in hand, we show how it applies to the
analysis of several CAs encountered in applications, and how it simplifies the
analysis of their particle dynamics.  We close with a summary of the results
and a list of open questions.  An appendix gives the details of the proof of an
auxiliary result on domains.

The present work is motivated by and bears on several larger issues. Of
particular relevance is the notion of an ``object'' or ``coherent structure''
that spontaneously emerges in the space-time behavior of a process
\cite{Anything-ever-new,Holland-emergence}. The particles analyzed here are
arguably one of the simplest kinds of such emergent structures. Despite this
interest, we do {\em not} define ``particle'' from first principles. Like our
predecessors we take the existence of particles as a given and assume we know
how to recognize them in the space-time behavior.  Nonetheless, the results and
their proofs do elucidate some of the component concepts that we feel will be
useful in a theory of emergent structures in spatial processes.\footnote{For
two approaches to the automatic discovery of particles, see
\cite{Wuesnche-glider-finding} and
\cite{Eppstein-searching-for-spaceships}.}

\section{Cellular Automata, Formal Languages, and Transducers}

A {\em cellular automaton} (CA) is a discrete dynamical system consisting of a
regular lattice of identical {\em cells}. At each time step $t$, each of these
cells is in one of a number $k$ of states $\Sigma$. The {\em state} of cell
$i$ at time $t$ is denoted $s^i_t \in \Sigma \equiv \{0,1,\ldots,k-1\}$. The
{\em global state}
${\bf s}_t$ of a one-dimensional CA at time $t$ is the {\em configuration} of
the entire lattice; i.e.,
${\bf s}_t = (s^0_t,s^1_t,\ldots,s^{N-1}_t) \in \Sigma^N$, where $N$ is the
lattice size. One often sees CA phenomenology studied where $\Sigma = \{0,1\}$
and with periodic boundary conditions: $s^{N+i}_t = s^i_t$. The main results
reported below do not depend on these restrictions, however.

At each next time step $t+1$, the cells in the lattice update their states
simultaneously according to a local update rule $\phi$. This {\em update rule}
$\phi$ takes as input the current local neighborhood configuration $\eta^i_t =
(s^{i-r}_t,\ldots,s^i_t,\ldots,s^{i+r}_t)$ of cell $i$ and returns the next
state $s^i_{t+1}$; $r$ is the CA's {\em radius}. Thus, the CA {\em equations of
motion} are given by
\begin{equation}
s^i_{t+1} = \phi(\eta^i_t) ~.
\end{equation}
The {\em global update rule} $\Phi : \Sigma^N \rightarrow \Sigma^N$ applies
$\phi$ in parallel (simultaneously) to all cells in the CA lattice; i.e.,
\begin{eqnarray}
{\bf s}_{t+1} & = & \Phi({\bf s}_t)\\
& = & \left(\phi(\eta^0_t),\phi(\eta^1_t), \ldots \phi(\eta^{N-1}_t)\right) ~.
\end{eqnarray}

Binary $(k=2)$ local state, $r=1$ CAs are referred to as {\em elementary} CAs (ECAs) \cite{Wolfram-stat-mech-CA}.

An ensemble operator $\boldsymbol{\Phi}$ can be defined
\cite{Attractor-basin-portrait,Wolfram-computation}
that operates on sets of lattice configurations $\Omega_t = \{{\bf s}_t\}$:
\begin{equation}
\Omega_{t+1} = \boldsymbol{\Phi}\Omega_t ~,
\end{equation}
such that
\begin{equation}
\Omega_{t+1} = \{ {\bf s}_{t+1} : {\bf s}_{t+1} = \Phi({\bf s}_t), ~
  {\bf s}_t \in \Omega_t \} ~.
\end{equation}

It is often informative to describe CA configurations as one or another type
of formal language. A {\em formal language} $\mathcal L$ over the alphabet $\Sigma$
is a subset of $\Sigma^*$---the set of all possible words, or strings, made up
of symbols from $\Sigma$. A {\em regular} language is a formal language whose words can
be generated or recognized by a device with finite memory; sometimes called
a {\em finite automaton}. Regular languages are the simplest class of formal
languages in a hierarchy (the Chomsky hierarchy) of language classes of
increasing complexity \cite{Hopcroft-Ullman}.

A {\em deterministic finite automaton} (DFA) $M$ is defined as a 5-tuple:
\begin{equation}
M = \{Q,\Sigma,\delta,q_0,F\} ~,
\end{equation}
where $Q$ is a finite set of {\em states}, $\Sigma$ is an {\em alphabet},
$q_0 \in Q$ is the {\em initial} state, $F \subseteq Q$ is a set of {\em final}
states, and $\delta : Q \times \Sigma \rightarrow Q$ is a
{\em transition function}: $\delta(q,a) = q'$, where
$q,q' \in Q$ and $a \in \Sigma$.

A DFA can be used to read, or scan, words $w = w_1 \ldots w_L$ over the
alphabet $\Sigma$. Starting in the initial state $q_0$, the DFA reads the
first symbol $w_1$ of the word $w$. It then makes a transition to another
state $q' = \delta(q_0,w_1)$. The DFA then reads the next symbol $w_2$ and
makes a transition to $q'' = \delta(q',w_2)$, and so on until all symbols in
$w$ have been read or until an undefined transition is encountered. If,
after reading $w$, the DFA ends in a final state $q \in F$, $M$ {\em accepts}
$w$; otherwise $M$ {\em rejects} it.

A regular language $\mathcal L$ is a formal language for which there exists a DFA
that accepts all words in $\mathcal L$ and rejects all words not in $\mathcal L$.  If
there is one such DFA, there are generally many of them, but there is a unique
minimal DFA for ${\mathcal L}$, which we write $M({\mathcal L})$.  Similarly, for every
DFA $M$ there is a corresponding regular language ${\mathcal L}(M)$ consisting of
all and only the words that are accepted by $M$. The {\em regular process
languages} are a subset of the regular languages: those containing all
subwords of words in the language. All states of the corresponding DFA---the
{\em process graph}---are both initial and accepting states
\cite{Attractor-basin-portrait}.

Finally, a {\em finite-state transducer} (FST) is a finite automaton with
two kinds of symbol associated with each transition: inputs and outputs.
Formally, an FST $R$ is defined as a 7-tuple:
\begin{equation}
R = \{Q,\Sigma_{in},\Sigma_{out},\delta,\lambda,q_0,F\} ~,
\end{equation}
where $Q,\delta,q_0,$ and $F$ are as in a DFA, $\Sigma_{in}$ is the 
{\em input alphabet}, $\Sigma_{out}$ is the {\em output alphabet}, and
$\lambda : Q \times \Sigma_{in} \rightarrow \Sigma_{out}$ is the
{\em observation function}:
$\lambda(q,a) = b$ where $q \in Q$, $a \in \Sigma_{in}$, and $b \in
\Sigma_{out}$. An FST effectively implements a mapping $f_R$ from one language
over $\Sigma_{in}$ to another language over $\Sigma_{out}$. In other words, it
reads a word $w \in \Sigma_{in}^*$ and transforms it to another word $w' \in
\Sigma_{out}^*$ by mapping each symbol $w_i \in \Sigma_{in}$ to a symbol $w_i'
\in \Sigma_{out}$ such that $w_i' = \lambda(q,w_i)$, where $q \in Q$ is the
current state of $R$ when reading $w_i$.

In formal language theory, languages and automata play the role of
sets and transducers the role of functions.

\section{Domains}

We are now ready to review the computational mechanics analysis of
emergent structures in CAs \cite{Attractor-basin-portrait,Inferring-stat-compl}.

A {\it regular domain} $\Lambda$ of a CA $\Phi$ is a process language,
representing a set of spatial lattice configurations, with the following
properties:
\begin{enumerate}
\item {\it Temporal invariance (or periodicity)}: $\Lambda$ is mapped onto itself
	by the CA dynamic; i.e., $\boldsymbol{\Phi}^p \Lambda = \Lambda$ for some
	finite $p$.  (Recall that $\mathbf{\Phi}$ takes sets of lattice
	configurations into sets of configurations and that a formal language,
	such as $\Lambda$, {\it is} a set of configurations.)
\item {\it Spatial homogeneity}: The {\em process graph} of each temporal
	iterate of $\Lambda$ is strongly connected. That is, there is a path
	between every pair of states in $M( \Phi^l \Lambda )$ for all $l$.
	(Recall that $M({\mathcal L})$ is the minimal DFA which recognizes the
	language $\mathcal L$.)
\end{enumerate}
The set of all domains of a CA $\Phi$ is denoted
$\boldsymbol{\Lambda} = \{ \Lambda^0, \Lambda^1, \dots, \Lambda^{m-1}\}$,
where $m = |\boldsymbol{\Lambda}|$.

According to the first property---temporal invariance or periodicity---a
particular domain $\Lambda^i$ consists of $p$ temporal phases for some $p \geq
1$; i.e., $\Lambda^i = \{ \Lambda_0^i, \Lambda_1^i, \ldots, \Lambda_{p-1}^i\}$,
such that $\boldsymbol{\Phi}^l \Lambda_j^i = \Lambda_{(j+l) \bmod p}^i$. Here
$p$ is the {\it temporal periodicity} of the domain $\Lambda^i$;
which we denote by $T(\Lambda^i)$.

Each of the temporal phases $\Lambda_j^i$ of a domain $\Lambda^i$ is
represented by a process graph $M(\Lambda^i_j)$ which, according to the second
property (spatial homogeneity), is strongly connected. Each of these process
graphs consists of a finite number of states. We denote the ${k}^{\mathrm th}$
state of the ${j}^{\mathrm th}$ phase of $\Lambda^i$ by $\Lambda_{j,k}^i$,
intentionally suppressing the $M( \cdot )$ notation for conciseness. We write
the number of states in a given phase as $S(\Lambda_j^i)$.

The process graphs of all temporal phases $\Lambda^i_j$ of all domains
$\Lambda^i$ can be connected together and transformed into a finite-state
transducer, called the {\it domain transducer}, that reads in a spatial
configuration and outputs various kinds of information about the sites.  (The
construction is given in, for example, \cite{Turbulent-pattern-bases}.)
Variations on this transducer can do useful recognition tasks. For example, all
transitions that were in domain $\Lambda^i_j$'s process graph are assigned
output symbol $D$, indicating that the input symbol being read is
``participating'' in a domain. All other transitions in the transducer indicate
deviations from the sites being in a domain. They can be assigned a unique
output (``wall'') symbol $w \in \{ W^i_j \}$ that labels the kind of domain
violation that has occurred. The resulting domain transducer can now be used to
{\em filter} CA lattice configuration, mapping all domain {\it regularities} to
$D$ and mapping all domain {\it violations} to output symbols $w$ that indicate
{\em domain walls} of various kinds.

We say that a phase of a domain is (spatially) {\em periodic} when the
process graph consists of a periodic chain of states, with a single transition
between successive states in the chain.  That is, as one moves from state to
state, an exactly periodic sequence of states is encountered and an exactly
periodic sequence of symbols from $\Sigma$ is encountered on the transitions.
The \textit{spatial periodicity} of a periodic phase is simply $S(\Lambda^i)$.
We say that a domain is periodic when all its phases are periodic.  Almost all
of our attention in the sequel will be confined to periodic domains, for the
following reason.  It turns out that for such domains all of the spatial
periodicities $S(\Lambda_j^i)$ at each temporal phase are equal. Thus, we can
speak of {\it the} spatial periodicity $S(\Lambda^i)$ of a periodic domain
$\Lambda^i$.  This property, in turn, is central to our proof of the upper
bound on the number of particle interaction products.

\begin{lemma}
If a domain $\Lambda^i$ has a periodic phase, then the domain is periodic, and
the spatial periodicities $S(\Lambda^i_j)$ of all its phases $\Lambda^i_j, j =
0, \ldots, p-1,$ are equal.
\label{PROPSPATIALPERIODICITY}
%\label{PropSpatialPeriodicity}
\end{lemma}

{\it Proof.} See the Appendix.\hfill$\Box$

Thus, the number of states in the process graph representing a particular
temporal phase $\Lambda^i_j$ is the same for all $j \in \{ 1, \ldots,
T(\Lambda^i) \}$, and it is, in fact, $S(\Lambda^i)$.

Finally, there is a larger class of {\em cyclic domains} whose process graphs
consist of a periodic chain of states: as one moves from state to state an
exactly periodic sequence of {\em states} is seen.  Note that this class
includes more than periodic domains, which are obviously cyclic. It includes
domains in which between two successive states in the chain there are multiple
transitions over $\Sigma$. (See \cite{Turbulent-pattern-bases} for a CA
exhibiting two such cyclic domains.) Based on our experience we conjecture that
Prop.~\ref{PROPSPATIALPERIODICITY} also holds for cyclic domains. If this is
so, most of the following results, and in particular the upper bound theorem,
would hold for this large class.

\begin{conjecture}
For any cyclic domain $\Lambda^i$, the spatial periodicities $S(\Lambda^i_j)$
of all its phases $\Lambda^i_j, j = 0, \ldots, p-1,$ are equal.
\end{conjecture}

\section{Particles}

When domain violations form a spatially localized (finite width), temporally
periodic boundary between two adjacent domains, they are called {\it
particles}.

\begin{definition}
A particle $\alpha$ is a set $\{ \alpha^0, \alpha^1, \ldots, \alpha^{p-1}\}$ of
finite-width words $\alpha^j$ over $\Sigma^*$, called {\rm wedges}, such that
\begin{equation}
{\boldsymbol \Phi} (\Lambda \alpha^{i} \Lambda^\prime)
  = \Lambda \alpha^{(i+1) \bmod p} \Lambda^\prime ~,
\label{PeriodicityCondition}
\end{equation}
for some finite $p$ and $\Lambda$ and $\Lambda^\prime \in \boldsymbol{\Lambda}$.
\end{definition}

Since a particle is a bounded structure, it does not have a spatial
periodicity.  In the following, the ``periodicity of a particle'' therefore
refers to temporal periodicity.

Since these particles are temporally periodic, we view the appearance of wedge
$\alpha^j$ as the particle being in it's $j$th {\em phase}. The $k$th symbol in
the wedge's word is denoted $\alpha^j_k$. The state in which the domain
transducer finds itself after reading the $k$th symbol $\alpha^j_k$ in the
wedge $\alpha^j$ is denoted $q(\alpha^j_k)$.

We now introduce an important but subtle distinction. The particle period $p$
referred to above---the {\em surface periodicity}---is associated with the
repetition over time of the wedge words as observed in the raw space-time
behavior ${\boldsymbol s}_0, {\boldsymbol s}_1, {\boldsymbol s}_2, \ldots$.  It
turns out, as will become clear, that particles have an internal periodicity
that may be some multiple of the surface periodicity $p$. The internal
periodicity---the one of actual interest here---though, is the periodicity seen
by the various phases of the bordering domains.

\begin{definition}
A particle $\alpha$'s {\rm intrinsic periodicity} $P(\alpha)$ is the
periodicity of the set of transducer-state sequences generated when reading
a particle's wedges. For wedge $\alpha^j = \alpha^j_0 \ldots \alpha^j_n$ the state sequence
$q(\alpha^j_0) \ldots q(\alpha^j_n)$ is generated in the transducer. We denote
this state sequence by $q(\alpha^j)$. $P(\alpha)$, then, is the number of
iterations over which the sequence $q(\alpha^j)$ reappears.
\end{definition}

{\em Remark 1.}
$P(\alpha)$ is an integer multiple of $\alpha$'s apparent periodicity.

{\em Remark 2.}
A simple illustration of the need for intrinsic, as opposed to merely surface,
periodicity is provided by the $\gamma$ particles of ECA 54.  See
Fig.~\ref{ECA54GammasFiltered}(b) and the accompanying text in
Sec.~\ref{subsection:ECA54}.

After one period $P(\alpha)$, a particle $\alpha$ will have moved a number
$d_\alpha$ of sites in the CA lattice. This shift $d_\alpha$ in space after
one period is called the particle's {\it displacement}.  $d_\alpha$ is
negative for displacements to the left and positive for displacements to the
right. From the particle's periodicity $P(\alpha)$ and displacement $d_\alpha$,
its average velocity is simply $v_\alpha = d_\alpha / P(\alpha)$.

Note that the above remarks hold whether we are looking at the wedges or at
the transducer-state labeled wedges: one obtains the same velocity.

The set of all particles $\alpha, \beta, \ldots$ of a CA $\Phi$
is denoted by $\bf P$.

{\it Remark 3.} Here we defined temporally periodic particles. There are
particles in CAs, such as in ECA 18, which are temporally aperiodic.
In this case, one replaces the periodicity condition Eq.
(\ref{PeriodicityCondition}) by one using the ensemble operator; viz.,
\begin{equation}  
\boldsymbol{\Phi}^p (\Lambda \alpha \Lambda^\prime)
  = \Lambda \alpha \Lambda^\prime ~.
\end{equation}

\subsection{Structural Complexity of a Particle}

The preceding definitions and discussion suggest that one can think of
particles as having an internal clock or, in the more general case that
includes aperiodic particles, an internal state, much as the solitary-wave
solutions of continuum envelope equations have internal states
\cite{Infeld-Rowlands}.  One can ask about how much information a particle
stores in its states. This is the amount of information that a particle
transports across space and time and brings to interactions. These
considerations lead one to a natural measure of the amount of structural
complexity associated with individual particles.

\begin{definition}
The {\rm structural complexity} $C(\alpha)$ of a particle $\alpha$ is defined
to be
\begin{equation}
C(\alpha) = - \sum_{j=0}^{p-1} {\rm Pr} (q(\alpha^j))
  \log_2{{\rm Pr} (q(\alpha^j))} ~,
\end{equation}
\end{definition}
where $p$ is $\alpha$'s period and ${\rm Pr} (q(\alpha^j))$ is the probability
of $\alpha$ being in phase $\alpha^j$ with the state-sequence $q(\alpha^j)$.

{\em Remark 1.} For the straightforward case of periodic particles, in which
the wedges and so their associated state sequences are equally probable,
we have
\begin{equation}
C(\alpha) = \log_2{P(\alpha)}~.
\end{equation}

{\em Remark 2.}
The information available to be processed in particle interactions is
upper-bounded by the sum of the individual particle complexities, since
this sum assumes independence of the particles. As we will see shortly,
the relative information---that information in one particle, conditioned
on the other's phase (via the constraints imposed by the mediating domain)
and suitably averaged---determines the information available for processing
by interactions.

\subsection{Domain Transducer View of Particle Phases}

A particle is bounded on either side by two patches of domain. (They could
be patches of the same or different domains.) Consider what happens to the
domain transducer as it scans across the part of the lattice containing the
bounding domains ($\Lambda^i$ and $\Lambda^{i^\prime}$) and the particle
($\alpha$).  It begins by cycling through the states of the process graph
of a phase ($j$) of the first bounding domain ($\Lambda^i$). It then
encounters a symbol that does not belong to the language of that domain phase,
and this then causes a transition out of that process graph. Each successive
symbol of the particle wedge leads to additional transitions in the transducer.
Finally, the transducer reaches cells at the beginning of the other bounding
domain ($\Lambda^{i^\prime}$), whereupon it begins to follow the process graph
of $\Lambda^{i^\prime}_{j^\prime}$ at some appropriate phase $j^\prime$. In
this way, a particle wedge $\alpha^j$ corresponds to a sequence $q(\alpha^j)$
of transducer states.

More formally, the transducer maps a particle wedge $\alpha^j$, bordered by
$\Lambda^i_j$ and $\Lambda^{i^\prime}_{j^\prime}$, to an ordered $n-$tuple
($n = |\alpha^j|+2$) of states
\begin{eqnarray}
Q(\alpha^j) & =  &
\left< q(\Lambda^i_{j,k}), q(\alpha^j),
  q(\Lambda^{i^\prime}_{j^\prime,k^\prime}) \right> ~,
\end{eqnarray}
where $q(\Lambda^i_{j,k})$ is the transducer state reach on
reading symbol $\Lambda^i_{j,k}$.
Since the transducer-state sequence is determined by the bounding domain
phases and the actual wedge $\alpha^j$, it follows that the mapping from
particle wedges to state sequences is 1-1. If two particle wedges correspond to
the same sequence of states, then they are the same phase of the same
particle, and vice versa.

This representation of particle phases will prove very handy below.

\section{Interactions}

In many CAs, when two or more particles collide they create another set of
particles or mutually annihilate. Such {\it particle interactions} are denoted
$\alpha + \beta \rightarrow \gamma$, for example. This means that the collision
of an $\alpha$ particle on the left and a $\beta$ particle on the right leads
to the creation of a $\gamma$ particle. Particle annihilation is denoted
$\alpha + \beta \rightarrow \emptyset$. For completeness we note that there
are also {\em unstable} walls that can spontaneously decay into particles.
This is denoted $\alpha \rightarrow \beta + \gamma$, for example.

Often, the actual product of a particle interaction depends on the phases
$\alpha^j$ and $\beta^k$ in which the interacting particles are at the time
of collision. In such a case, there can be more than one interaction product
for a particular collision: e.g., both $\alpha + \beta \rightarrow \gamma$
and $\alpha + \beta \rightarrow \emptyset$ can be observed.

The set of a CA's possible particle interactions is denoted $\bf I$. The
complete information about a CA's domains $\boldsymbol{\Lambda}$, particles
$\bf P$, and particle interactions $\bf I$ can be summarized in a {\it particle
catalog}. The catalog forms a high-level description of the CA's dynamics. It
is high-level in the sense of capturing the dynamics of emergent structures.
The latter are objects on a more abstract level than the original equations
of motion and raw (uninterpreted) spatial configurations of site values.

\section{Bounding the Number of Interaction Products}

\begin{figure}
\epsfxsize=3.5in
\begin{center}
\leavevmode
\epsffile{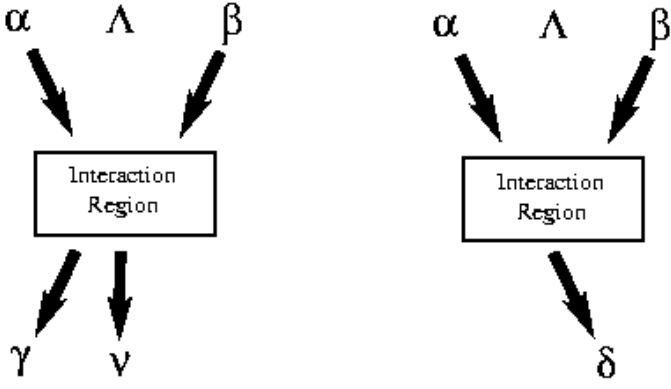}
\end{center}
\caption{Interactions between an $\alpha$ and a $\beta$ particle with
  domain $\Lambda$ lying between.
  }
\label{InteractionRegion}
\end{figure}

Restricting ourselves to particle interactions with just two colliding
particles---$\alpha$ and $\beta$, say---we now give an upper bound on the
number $n_{\alpha,\beta}$ of possible interaction products from a collision
between them. (See Fig.~\ref{InteractionRegion} for the interaction
geometry.) In terms of the quantities
just defined, the upper bound, stated as Thm.~\ref{ThmUpperBound} below, is:
\begin{equation}
n_{\alpha,\beta} \leq
	\frac{P(\alpha) P(\beta) \Delta v}{T(\Lambda^i) S(\Lambda^i)} ~,
\label{UpperBound}
\end{equation}
where $\Delta v = v_\alpha - v_\beta > 0$ and $\Lambda^i$ is the domain in
between the two particles before they collide.
Note that if $\Delta v = 0$, then $n_{\alpha,\beta} = 0$ trivially.

For simplicity, in the rest of the development we assume that
$\Delta v = v_{\alpha} - v_{\beta} \geq 0$. This simply means that particle
$\alpha$ lies to the left of $\beta$ and they move closer to each other over
time, as in Fig. \ref{InteractionRegion}.

This section proves that Eq. (\ref{UpperBound}) is indeed a proper upper
bound. The next section gives a number of examples, of both simple and
complicated CAs, that show the bound is and is not attained. These highlight
an important distinction between the number of possible interactions (i.e.,
what can enter the interaction region) and the number of unique interaction
products (i.e., what actually leaves the interaction region).

To establish the bound, we collect several intermediate facts. The first
three lemmas come from elementary number theory. Recall that the {\it least
common multiple} $\lcm (a,b)$ of two integers $a$ and $b$ is the
smallest number $c$ that is a multiple of both $a$ and $b$. Similarly, the
{\it greatest common divisor} $\gcd (a,b)$ of two integers $a$ and $b$
is the largest number $c$ that divides both $a$ and $b$.

\begin{lemma}
$\gcd(ca,cb) = c \; \gcd(a,b),\;c > 0$.
\label{lemma:cgcd}
\end{lemma}
{\it Proof.} See Thm. 2.7 in \cite{Burton-number-theory}.\hfill$\Box$

\begin{lemma}
$\gcd(a,b) \; \lcm(a,b) = ab$.
\label{lemma:gcdlcm}
\end{lemma}
{\it Proof.} See Thm. 2.8 in \cite{Burton-number-theory}.\hfill$\Box$

\begin{lemma}
$\lcm(ca,cb) = c \; \lcm(a,b),\;c > 0$.
\label{lemma:clcm}
\end{lemma}
{\it Proof.} Using Lemmas \ref{lemma:cgcd} and \ref{lemma:gcdlcm}, it
follows that
\begin{eqnarray}
\lcm(ca,cb) & = & \frac{cacb}{\gcd(ca,cb)} \\ \nonumber
				  & = & c \frac{ab}{\gcd(a,b)} \\ \nonumber
                  & = & c \; \lcm(a,b) ~.
\end{eqnarray}
\hfill$\Box$

Now we are ready to begin building an analysis of particles and particle
interactions.

\begin{lemma}
The intrinsic periodicity $P(\alpha)$ of a particle $\alpha$ is a multiple
of the temporal periodicity $T(\Lambda^i)$ of either domain $\Lambda^i$
for which $\alpha$ is a boundary. That is,
\begin{equation}
P(\alpha) = m_{\alpha i} T(\Lambda^i) ~,
\end{equation}
for some positive integer $m_{\alpha i}$ that depends on $\alpha$ and
$\Lambda^i$.
\label{lemma:mp_lambda}
\end{lemma}
{\it Proof.}
At any given time, a configuration containing the particle $\alpha$ consists of
a patch of the domain $\Lambda^i$, a wedge belonging to $\alpha$, and
then a patch of $\Lambda^{i^\prime}$, in that order from left to right.  (Or
right to left, if that is the chosen scan direction.)  Fix the phase of $\alpha$ to be
whatever we like --- $\alpha^l$, say.  This determines the phases of
$\Lambda^i$, for the following reason. Recall that, being a phase of a
particle, $\alpha^l$ corresponds to a unique sequence $Q(\alpha^l)$ of
transitions in the domain transducer. That sequence starts in a particular
domain-phase state $\Lambda^i_{j, k}$ and ends in another domain-phase state
$\Lambda^{i^\prime}_{j^\prime, k^\prime}$. So, the particle phase $\alpha^l$
occurs only at those times when $\Lambda^i$ is in its $j^{\rm th}$ phase. Thus,
the temporal periodicity of $\alpha$ must be an integer multiple of the
temporal periodicity of $\Lambda^i$. By symmetry, the same is also true for the
domain $\Lambda^{{i}^{\prime}}$ to the right of the wedge.\hfill$\Box$

\begin{corollary}
Given that the domain $\Lambda^i$ is in phase $\Lambda^i_j$ at some time step,
a particle $\alpha$ forming a boundary of $\Lambda^i$ can only be in a fraction
$1/T(\Lambda^i)$ of its $P(\alpha)$ phases at that time.
\label{cor:phase-restriction}
\end{corollary}

{\it Proof.} This follows directly from Lemma \ref{lemma:mp_lambda}.

{\it Remark.}  Here is the first part of the promised restriction on the
information in multiple particles.  Consider two particles $\alpha$ and
$\beta$, separated by a domain $\Lambda^0$.  Naively, we expect $\alpha$ to
contain $\log_2{P(\alpha)}$ bits of information and $\beta$, $\log_2{P(\beta)}$
bits.  Given the phase of $\alpha$, however, the phase of $\Lambda^0$ is fixed,
and therefore the number of possible phases for $\beta$ is reduced by a factor
of $1/T(\Lambda^0)$.  Thus the number of bits of information in the
$\alpha$-$\beta$ pair is at most
\begin{equation}
\log_2{P(\alpha)} + \log_2{P(\beta)} - \log_2{T(\Lambda^0)} =
\log_2{\frac{P(\alpha)P(\beta)}{T(\Lambda^0)}} ~.
\end{equation}
The argument works equally well starting from $\beta$.

\begin{lemma}
For any two particles $\alpha$ and $\beta$, the quantity $\lcm(P(\alpha),P(\beta))\Delta v$ is a non-negative integer.
\label{lemma:integer-displacement}
\end{lemma}
{\it Proof.}  We know that the quantity is non-negative, since the least common
multiple always is and $\Delta v$ is so by construction. It remains to show
that their product is an integer.  Let ${k_\alpha} = {\lcm(P(\alpha),P(\beta))
/ P(\alpha)}$ and ${k_\beta} = {\lcm(P(\alpha),P(\beta)) / P(\beta)}$; these
are integers.  Then
\begin{eqnarray*}
\Delta v & \equiv & \frac{d_\alpha}{P(\alpha)} - \frac{d_\beta}{P(\beta)} \\
& = & \frac{{k_\alpha}{d_\alpha} - {k_\beta}{d_\beta}}{\lcm(P(\alpha),P(\beta))} ~.
\end{eqnarray*}
When multiplied by $\lcm(P(\alpha),P(\beta))$ this is just
${k_\alpha}{d_\alpha} - {k_\beta}{d_\beta}$, which is an integer.
\hfill$\Box$
\begin{lemma}
When the distance $d$ between two approaching particles $\alpha$ and $\beta$,
in phases $\alpha^j$ and $\beta^{j^\prime}$, respectively, is increased by
$\lcm(P(\alpha), P(\beta))\Delta v$ sites, the original
configuration---distance $d$ and phases $\alpha^j$ and $\beta^{j^\prime}$---recurs
after $\lcm (P(\alpha), P(\beta))$ time steps.
\label{lemma:space_shift}
\end{lemma}
{\it Proof.} From the definition of $\lcm(a,b)$ it follows directly that
$\lcm(P(\alpha),P(\beta))$ is a multiple of $P(\alpha)$.  Thus,
\begin{equation}
\alpha^{(j+ \lcm(P(\alpha),P(\beta))) \bmod P(\alpha)} = \alpha^j ~,
\end{equation}
and the $\alpha$
particle has returned to its original phase.  Exactly parallel reasoning holds
for the $\beta$ particle.  So, after $\lcm(P(\alpha), P(\beta))$ time steps
both $\alpha$ and $\beta$ are in the same phases $\alpha^j$ and
$\beta^{j^\prime}$ again. Furthermore, in the same amount of time the distance
between the two particles has decreased by $\lcm(p_\alpha,p_\beta)\Delta v$,
which is the amount by which the original distance $d$ was increased. (By Lemma
\ref{lemma:integer-displacement}, that distance is an integer, and so we can
meaningfully increase the particles' separation by this amount.)  Thus, after
$\lcm( P(\alpha), P(\beta) )$ time steps the original configuration is
restored.\hfill$\Box$

\begin{lemma}
If $\Lambda^i$ is the domain lying between two particles $\alpha$ and $\beta$,
then the ratio
\begin{equation}
r = \frac{ \lcm(P(\alpha),P(\beta))\Delta v }{ S(\Lambda^i)}
\end{equation}
is an integer.
\label{lemma:integer_value}
\end{lemma}
{\it Proof.} Suppose, without loss of generality, that the particles begin in
phases $\alpha^0$ and $\beta^0$, at some substantial distance from each other.
We know from the previous lemma that after a time $\lcm(P(\alpha),P(\beta))$
they will have returned to those phases and narrowed the distance between each
other by $\lcm(P(\alpha),P(\beta))\Delta v$ cells.  What the lemma asserts is
that this displacement is some integer multiple of the spatial periodicity of
the intervening domain $\Lambda^i$.  Call the final distance between the
particles $d$. Note that the following does not depend on what $d$ happens to
be.

Each phase of each particle corresponds to a particular sequence of transducer
states---those associated with reading the particle's wedge for that phase.
Reading this wedge from left to right (say), we know that 
$Q(\alpha^0)$ must end in some phase-state of the domain $\Lambda^i$; call it
$\Lambda^i_{0, 0}$. Similarly, $Q(\beta^0)$
must {\it begin} with a phase-state of $\Lambda^i$, but, since every part
of the intervening domain is in the same phase, this must be a state of
the {\it same} phase $\Lambda^i_0$; call it $\Lambda^i_{0, k}$. In particular,
consistency requires that $k$ be the distance between the particles modulo
$S(\Lambda^i)$.  But this is true both in the final configuration, when the
separation between the particles is $d$, and in the initial configuration, when
it is $d + \lcm(P(\alpha),P(\beta))\Delta v $.  Therefore
\begin{eqnarray*}
d + \lcm(P(\alpha),P(\beta))\Delta v & = & d \pmod{S(\Lambda^i)} \\
\lcm(P(\alpha),P(\beta))\Delta v  & = & 0 \pmod{S(\Lambda^i)}~.
\end{eqnarray*}
Thus, $\lcm(P(\alpha),P(\beta))\Delta v$ is an integer multiple of the spatial
period $S(\Lambda^i)$ of the intervening domain $\Lambda^i$.\hfill$\Box$

{\it Remark.}  It is possible that $\lcm(P(\alpha),P(\beta))\Delta v = 0$,
but this does not affect the subsequent argument.  Note that if this is the
case, then, since the least common multiple of the periods is at least $1$,
we have $\Delta v = 0$. This, in turn, implies that the particles do not,
in fact, collide and interact, and so the number of interaction products is
simply zero. The formula gives the proper result in this case.

The next result follows easily from Lemmas \ref{lemma:cgcd} and
\ref{lemma:mp_lambda}.

\begin{lemma}
If $\Lambda^i$ is the domain lying between particles $\alpha$ and $\beta$, then
\begin{equation}
\gcd(P(\alpha), P(\beta))
  = T(\Lambda^i) \gcd(m_{\alpha i},m_{\beta i}) ~.
\end{equation}
\label{lemma:p_lambdagcd}
\end{lemma}
{\it Proof.}
We apply Lemma \ref{lemma:clcm}:
\begin{eqnarray*}
\gcd(P(\alpha), P(\beta))
&   = & \gcd(m_{\alpha i}T(\Lambda^i), m_{\beta i}T(\Lambda^i) \\
&  =  & T(\Lambda^i) \gcd(m_{\alpha i}, m_{\beta i}). 
\end{eqnarray*}

\hfill$\Box$

With the above lemmas the following theorem can be proved, establishing
an upper bound on the number of possible particle interaction products.

\begin{theorem}
The number $n_{\alpha,\beta}$ of products of an interaction between two
approaching particles $\alpha$ and $\beta$ with a domain $\Lambda^i$
lying between is at most
\begin{equation}
n_{\alpha,\beta} \leq
  \frac{P(\alpha) P(\beta) \Delta v} {T(\Lambda^i) S(\Lambda^i)} ~.
\end{equation}
\label{ThmUpperBound}
\end{theorem}
{\it Proof.} First, we show that this quantity is an integer. We use Lemma 
\ref{lemma:gcdlcm} to note that
\begin{equation}
\frac{P(\alpha) P(\beta)\Delta v}{T(\Lambda^i) S(\Lambda^i)}
 = \frac{\gcd(P(\alpha), P(\beta))\lcm(P(\alpha), P(\beta))\Delta v}{T(\Lambda^i) S(\Lambda^i)} ~,
\end{equation}
and then Lemma \ref{lemma:integer_value} to find that
\begin{equation}
\frac{P(\alpha) P(\beta)\Delta v}{T(\Lambda^i) S(\Lambda^i)}
= \frac{\gcd(P(\alpha), P(\beta))r}{T(\Lambda^i)} ~,
\end{equation}
and finally Lemma \ref{lemma:p_lambdagcd} to show that
\begin{eqnarray}
\nonumber
\frac{P(\alpha) P(\beta)\Delta v}{T(\Lambda^i) S(\Lambda^i)}
 & = & \frac{T(\Lambda^i) \gcd(m_{\alpha i},m_{\beta i}) r}{T(\Lambda^i)} \\
& = &r \gcd(m_{\alpha i},m_{\beta i}) ~,
\end{eqnarray}
which is an integer.

Second, assume that, at some initial time $t$, the two particles are in some
arbitrary phases $\alpha^j$ and $\beta^{j^\prime}$, respectively, and that the
distance between them is $d$ cells. This configuration gives rise to a
particular particle-phase combination at the time of collision. Since the
global update function is deterministic, the combination, in turn, gives one
and only one interaction result. Now, increase the distance between the two
particles, at time $t$, by one cell, while keeping their phases fixed. This
gives rise to a different particle-phase combination at the time of collision
and, thus, possibly to a different interaction result.  We can repeat this
operation of increasing the distance by one cell
$\lcm(P(\alpha),P(\beta))\Delta v$ times. At that point, however, we know from
Lemma $\ref{lemma:space_shift}$ that after $\lcm(P(\alpha),P(\beta))$ time
steps the particles find themselves again in phases
$\alpha^j$ and $\beta^{j^\prime}$ at
a separation of $d$. That is, they are in exactly the original configuration
and their interaction will therefore also produce the original product,
whatever it was.

Starting the two particles in phases $\alpha^j$ and $\beta^{j^\prime}$, the
particles go through a fraction $1 / \gcd(P(\alpha), P(\beta))$ of the possible
$P(\alpha)P(\beta)$ phase combinations, over $\lcm(p_{\alpha},p_{\beta})$ time
steps, before they start repeating their phases again. So, the operation of
increasing the distance between the two particles by one cell at a time needs
to be repeated for $\gcd(P(\alpha), P(\beta))$ different initial phase
combinations. This way all possible phase combinations with all possible
distances (modulo $\lcm(P(\alpha),P(\beta))\Delta v$) are encountered. Each of
these can give rise to a different interaction result.

From this one sees that there are at most
\begin{equation}
\gcd(P(\alpha), P(\beta))\lcm(P(\alpha), P(\beta)) \Delta v
  = P(\alpha)P(\beta)\Delta v
\end{equation}
unique particle-domain-particle configurations. And so, there are at most this
many different particle interaction products, given that $\Phi$ is many-to-one.
(Restricted to the homogeneous, quiescent ($\Lambda = 0^*$) domain which has
$T(\Lambda) = 1$ and $S(\Lambda) = 1$, this is the result, though not the
argument, of \cite{Park-Steiglitz-Thurston-soliton}.)

However, given the phases $\alpha^j$ and $\beta^{j^\prime}$, the distance
between the two particles cannot always be increased by an arbitrary number of
cells.  Keeping the particle phases $\alpha^j$ and $\beta^{j^\prime}$ fixed,
the amount $\Delta d$ by which the distance between the two particles can be
increased or decreased is a multiple of the spatial periodicity $S(\Lambda^i)$
of the intervening domain. The argument for this is similar to that in the
proof of Lemma \ref{lemma:integer_value}. Consequently, of the $\lcm(P(\alpha),
P(\beta)) \Delta v$ increases in distance between the two particles, only a
fraction $1 / S(\Lambda^i)$ are actually possible.

Furthermore, and similarly, not all arbitrary particle-phase combinations are
allowed. Choosing a phase $\alpha^j$ for the $\alpha$ particle subsequently
determines the phase $\Lambda^i_j$ of the domain $\Lambda^i$ for which $\alpha$
forms one boundary. From Corollary \ref{cor:phase-restriction} it then follows
that only a fraction $1 / T(\Lambda^i)$ of the $P(\beta)$ phases are possible
for the $\beta$ particle which forms the other boundary of $\Lambda^i$.

Adjusting the number of possible particle-domain-particle configurations
that can give rise to different interaction products according to the above
two observations results in a total number
\begin{equation}
\frac{P(\alpha) P(\beta) \Delta v} {T(\Lambda^i) S(\Lambda^i)}
\end{equation}
of different particle-phase combinations and distances between two particles
$\alpha$ and $\beta$. Putting the pieces together, then, this number is an
upper bound on the number $n_{\alpha,\beta}$ of different interaction products.
\hfill$\Box$

{\it Remark 1.}  As we shall see in the examples, on the one hand, the upper
bound is strict, since it is saturated by some interactions. On the other
hand, there are also interactions that do not saturate it.

{\it Remark 2.} We have seen (Corollary \ref{cor:phase-restriction}, Remark)
that the information in a pair of particles $\alpha$ and $\beta$, separated by
a patch of domain $\Lambda^i$, is at most
\begin{equation}
\log_2{\frac{P(\alpha)P(\beta)}{T(\Lambda^i)}}
\end{equation}
bits.  In fact, the theorem implies a stronger restriction.  The amount of
information the interaction carries about its inputs is, at most,
$\log_2{n_{\alpha,\beta}}$ bits, since there are only $n_{\alpha,\beta}$
configurations of the particles that can lead to distinct outcomes.  If the
number of outcomes is less than $n_{\alpha,\beta}$, the interaction effectively
performs an irreversible logical operation on the information contained in the
input particle phases.

\section{Examples}

\subsection{ECA 54 and Intrinsic Periodicity}
\label{subsection:ECA54}

\begin{figure}
\begin{center}
\epsfig{file=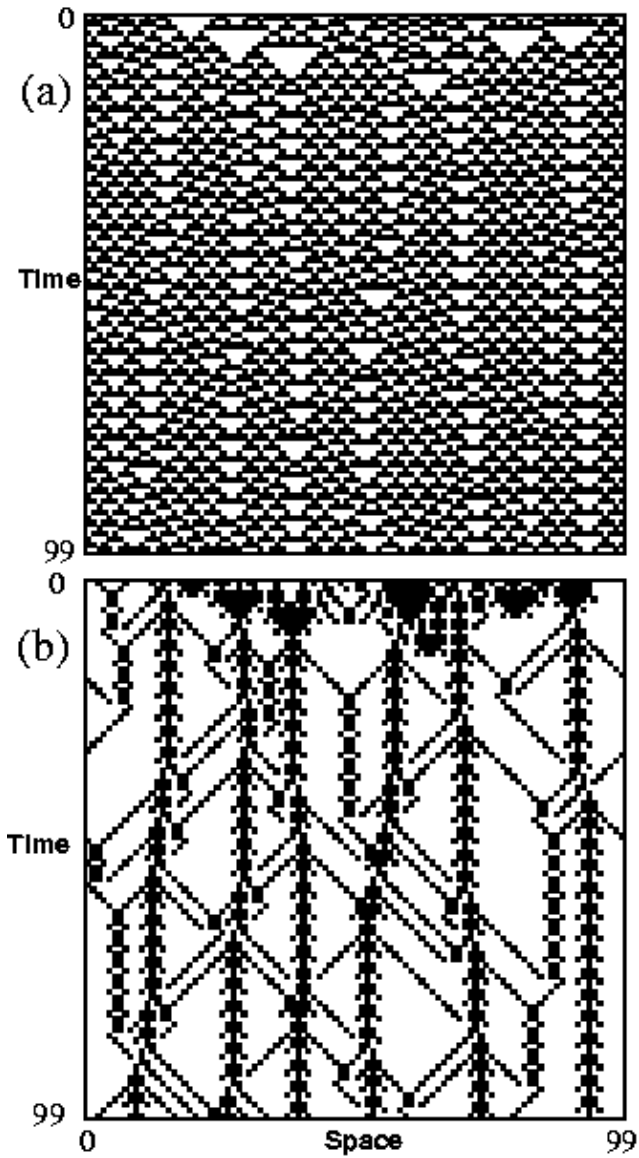,width=3.3in}
\end{center}
\caption{(a) Raw space-time diagram and (b) filtered space-time diagram of
  ECA 54 behavior starting from an arbitrary initial configuration.
  After \protect\cite{Comp-mech-of-CA-example}.
  }
\label{ECA54SpTmRawFiltered}
\end{figure}

Figure~\ref{ECA54SpTmRawFiltered} shows the raw and domain-transducer filtered
space-time diagrams of ECA 54, starting from a random initial configuration.
We first review the results of \cite{Comp-mech-of-CA-example} for ECA 54's
particle dynamics.

\begin{figure}
\begin{center}
\epsfig{file=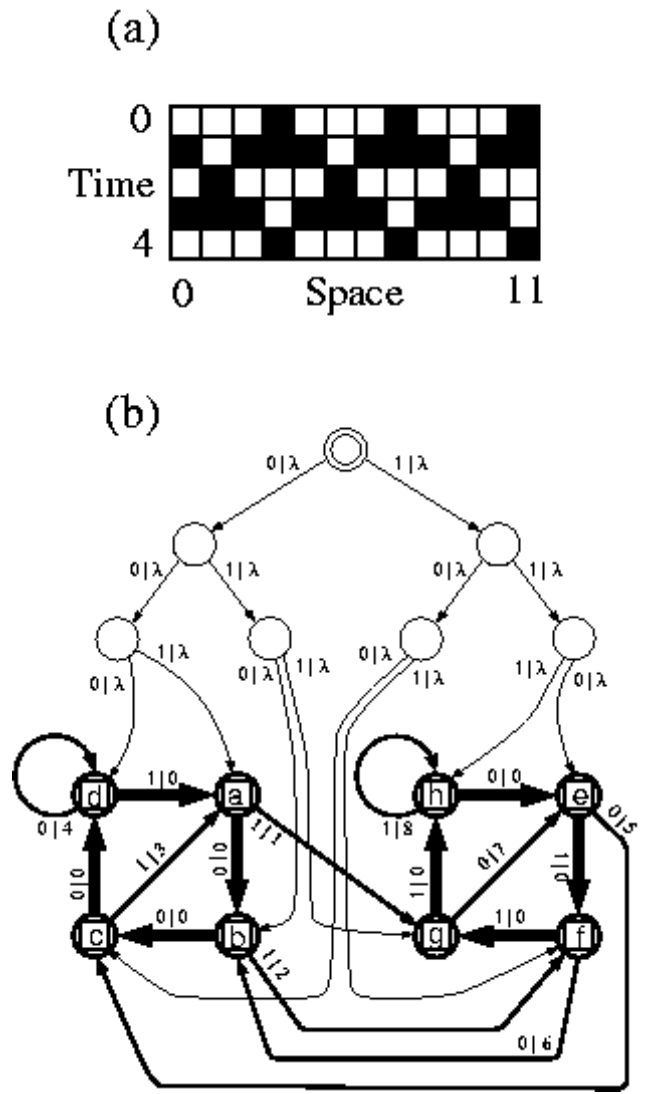,width=3.3in}
\end{center}
\caption{(a) Space-Time patch of ECA54's primary domain $\Lambda$.
  (b) The transducer that recognizes $\Lambda$ and deviations from it.
  After \protect\cite{Comp-mech-of-CA-example}.
  }
\label{ECA54DomainFilter}
\end{figure}

Figure~\ref{ECA54DomainFilter} shows a space-time patch of ECA 54's
dominant domain $\Lambda$, along with the domain transducer constructed
to recognize and filter it out, as was done to produce
Fig.~\ref{ECA54SpTmRawFiltered}(b).

Examining Fig. \ref{ECA54SpTmRawFiltered} shows that there are four particles;
we label these $\alpha$, $\beta$, $\gamma^+$, and $\gamma^-$. The first two
have zero velocity; they are the larger particles seen in
Fig. \ref{ECA54SpTmRawFiltered}(b). The $\gamma$ particles have velocities $1$
and $-1$, respectively. They are seen in the figure as the diagonally moving
``light'' particles that mediate between the ``heavy'' $\alpha$ and $\beta$
particles.

The analysis in \cite{Comp-mech-of-CA-example} identified $7$ dominant two- and
three-particle interactions. We now analyze just one: the $\gamma^+ + \gamma^-
\rightarrow \beta$ interaction to illustrate the importance of a particle's
intrinsic periodicity.

Naive analysis would simply look at the space-time diagram, either the raw or
filtered ones in Fig. \ref{ECA54SpTmRawFiltered}, and conclude that these
particles had periodicities $P(\gamma^+) = P(\gamma^-) = 1$.  Plugging this and
the other data---$T(\Lambda) = 2$, $S(\Lambda) = 4$, and $\Delta v =
2$---leads to upper bound $n_{\alpha,\beta} = 1/4$!  This is patently wrong;
it's not even an integer.

\begin{figure}
\begin{center}
\epsfig{file=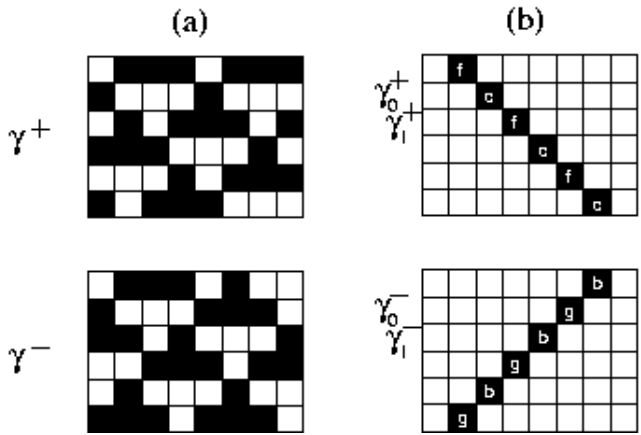,width=3.3in}
\end{center}
\caption{The transducer-filtered space-time diagrams for the $\gamma^+$
  and $\gamma^-$ particles. (a) The raw space-time patches containing
  the particles. (b) The same patches with the $\Lambda$ filtered out.
  The cells not in $\Lambda$ are denoted in black; those in $\Lambda$
  in white. In the filtered diagrams the transducer state reached on
  scanning the particle wedge cells is indicated.
  After \protect\cite{Comp-mech-of-CA-example}.
  }
\label{ECA54GammasFiltered}
\end{figure}

Figure~\ref{ECA54GammasFiltered} gives the transducer-filtered space-time
diagram for the $\gamma^+$ and $\gamma^-$ particles. The domain $\Lambda$ is
filtered out, as above. In the filtered diagrams the transducer state reached
on scanning the particle wedge cells is indicated.

From the space-time diagrams of Fig. \ref{ECA54GammasFiltered}(b) one notes that
the transducer-state labeled wedges for each particle indicate that their
intrinsic periodicities are $P(\gamma^+) = 2$ and $P(\gamma^-) = 2$. Then, from
Thm.~\ref{ThmUpperBound} we have that $n_{\alpha,\beta} = 1$. That is, there is
at most one product of these particles' interaction.

\begin{figure}
\begin{center}
\epsfig{file=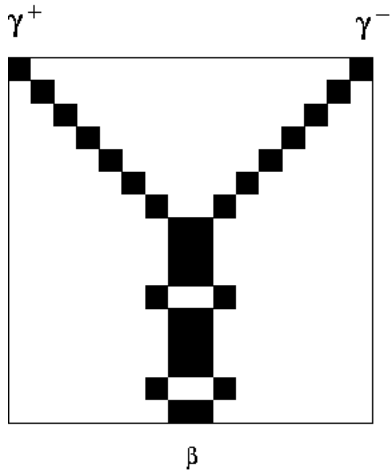,width=2.0in}
\end{center}
\caption{The transducer-filtered space-time diagrams for the
  $\gamma^+ + \gamma^- \rightarrow \beta$ interaction.
  After \protect\cite{Comp-mech-of-CA-example}.
  }
\label{ECA54GammaInteraction}
\end{figure}

Fig.~\ref{ECA54GammaInteraction} gives the transducer-filtered space-time
diagram for the $\gamma^+ + \gamma^- \rightarrow \beta$ interaction.  A
complete survey of all possible $\gamma^+$-$\Lambda$-$\gamma^-$ initial
particle configurations shows that this is the only interaction for these
particles. Thus, the upper bound is saturated.

\subsection{An Evolved CA}

The second example for which we test the upper bound is a CA that was
evolved by a genetic algorithm to perform a class of spatial
computations: from all random initial configurations, synchronize
within a specified number of iterations. This CA is $\phi_{sync_1}$
of \cite{Wim-MM-JPC-mechanisms}: a binary, radius-$3$ CA. The
$128$-bit look-up table for $\phi_{sync_1}$ is given
in Table \ref{Table:PhiLUTs}.

Here we are only interested in locally analyzing the various pairwise particle
interactions observed in $\phi_{sync_1}$. It turned out that this CA used a
relatively simple set of domains, particles, and interactions. Its particle
catalog is given in Table \ref{table:sync1_catalog}.

As one example, the two particles $\alpha$ and $\beta$ and the
intervening domain $\Lambda$ have the properties given in Table
\ref{table:sync1_catalog}. From this data, Thm.~\ref{ThmUpperBound}
tells us that there is at most one interaction product:
\begin{equation}
n_{\alpha,\beta} = \frac{4 \cdot 2 \cdot \frac{1}{4}}{2 \cdot 1} = 1 ~.
\end{equation}

\begin{table}
\begin{center}
\begin{tabular}{cc}
  $\phi$ & Look-up Table (hexadecimal) \\
\hline
  $\phi_{sync_1}$ & {\tt F8A19CE6B65848EA} \\
                  & {\tt D26CB24AEB51C4A0} \\
\hline
  $\phi_{parent}$ & {\tt CEB2EF28C68D2A04} \\
                  & {\tt E341FAE2E7187AE8} \\
\end{tabular}
\end{center}
\caption{Lookup tables (in hexadecimal) for $\phi_{sync_1}$ and
  $\phi_{parent}$. To recover the 128-bit string giving the CA look-up
  table output bits $s_{t+1}$, expand each hexadecimal digit (the
  first row followed by the second row) to binary.  The output bits
  $s_{t+1}$ are then given in lexicographic order starting from the
  all-$0$s neighborhood at the leftmost bit in the 128-bit
  string.}
  \label{Table:PhiLUTs}
\end{table}

%\medskip

% LUT Phi_sync_1 binary:
% 11111000101000011001110011100110101101100101100001001000111010101101001001101100101100100100101011101011010100011100010010100000
% LUT Phi_sync_1 hex:
% F8A19CE6B65848EAD26CB24AEB51C4A0

\begin{figure}
\begin{center}
\epsfig{file=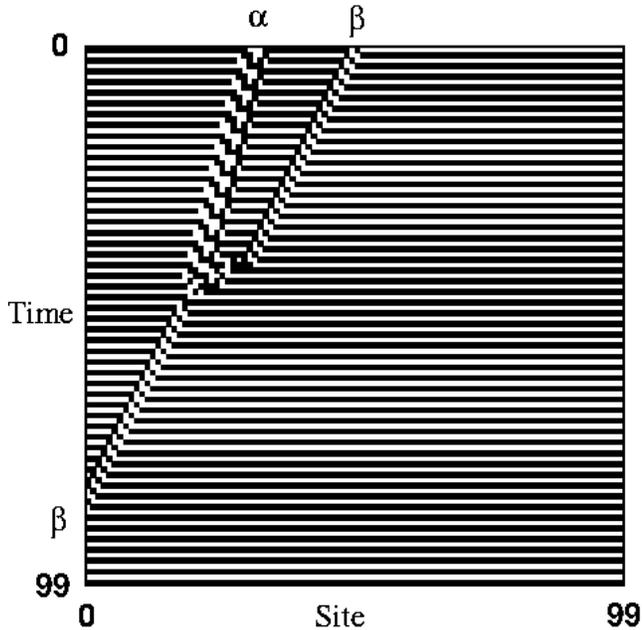,width=3.3in}
\end{center}
\caption{The interaction between an $\alpha$ and a $\beta$ particle
  in $\phi_{sync_1}$.}
\label{figure:example1}
\end{figure}

The single observed interaction between the $\alpha$ and $\beta$ particles
is shown in Fig. \ref{figure:example1}. As this space-time
diagram shows, the interaction creates another $\beta$ particle, i.e.,
$\alpha+\beta \rightarrow \beta$. An exhaustive survey of the $8$
($= 4 \times 2$) possible particle-phase configurations shows that this is
the only interaction for these two particles. Thus, in this case, we see that
Thm.~\ref{ThmUpperBound} again gives a tight bound; it cannot be reduced.

%%
%% Particle catalog of phi_sync_1.
%%
\begin{table}
  \begin{center}
  \begin{tabular}{ccccc}
  \multicolumn{5}{c}{\bf $\phi_{sync_1}$ Particle Catalog} \\
  \hline
  \multicolumn{5}{c}{\bf Domains ${\boldsymbol{\Lambda}}$} \\
  \hline
  Name & \multicolumn{2}{c}{Regular language} & $T(\Lambda)$ & $S(\Lambda)$ \\
  \hline
  $\Lambda$    & \multicolumn{2}{c}{$0^40^*$, $1^41^*$} & 2 & 1 \\
  \hline
  \multicolumn{5}{c}{\bf Particles P} \\
  \hline
  Name     & Wall                         & $P$ & $d$ & $v$ \\
  \hline
  $\alpha$ & $\Lambda_j \Lambda_j$     &   4 &  -1 &-1/4 \\
  $\beta$  & $\Lambda_j \Lambda_{1-j}$ &   2 &  -1 &-1/2 \\
  $\gamma$ & $\Lambda_j \Lambda_j$     &   8 &  -1 &-1/8 \\
  $\delta$ & $\Lambda_j \Lambda_j$     &   2 &   0 &   0 \\
  \hline
  \multicolumn{5}{c}{\bf Interactions I} \\
  \hline
  Type & \multicolumn{2}{c}{Interaction} & \multicolumn{2}{c}{Interaction} \\
  \hline
  React & \multicolumn{2}{c}{$\alpha + \beta \rightarrow \beta$}
    & \multicolumn{2}{c}{$\gamma+\beta \rightarrow \beta$} \\
  React & \multicolumn{2}{c}{$\delta+\beta \rightarrow \beta$}
    & \multicolumn{2}{c}{$\gamma+\alpha \rightarrow \alpha$} \\
  React & \multicolumn{2}{c}{$\delta+\alpha \rightarrow \alpha$}
    & \multicolumn{2}{c}{$\delta+\gamma \rightarrow \alpha$} \\
  \end{tabular}
  \end{center}
  \caption{The particle catalog of $\phi_{sync_1}$. $\Lambda_j$,
  $j \in \{0,1\}$, indicates the two temporal phases of domain $\Lambda$.}
  \label{table:sync1_catalog}
\end{table}

\subsection{Another Evolved CA}

The third, more complicated example is also a CA that was evolved by a
genetic algorithm to synchronize. This CA is $\phi_{parent}$ of
\cite{JPC-Wim-MM-performance}. It too is a binary radius-3 CA. The
$128$-bit look-up table for $\phi_{parent}$ was given in Table
\ref{Table:PhiLUTs}.

% LUT phi_parent binary:
%11001110101100101110111100101000110001101000110100101010000001001110001101000001111110101110001011100111000110000111101011101000
% LUT phi_parent hex:
% CEB2EF28C68D2A04E341FAE2E7187AE8

Here the two particles $\alpha$ and $\beta$ and the intervening domain
$\Lambda$ have the properties given in Table \ref{table:PhiParentProperties}.
Note that this is the same domain as in the preceding example.

\begin{table}
  \begin{center} 
  \begin{tabular}{cccc} 
  \multicolumn{4}{c}{\bf $\phi_{parent}$ Particle Properties} \\
  \hline
  Domain    & $T$ & $S$ & \\ 
  \hline 
  $\Lambda$ &  2  &  1  & \\ 
  \hline
  Particle  & $P$ & $d$ &  $v$ \\ 
  \hline 
  $\alpha$  &  8  &  2  &  1/4 \\ 
  \hline 
  $\beta$   &  2  & -3  & -3/2 \\ 
  \end{tabular} 
  \end{center} 
  \caption{Properties of two of $\phi_{parent}$'s particles.} 
  \label{table:PhiParentProperties} 
\end{table}

From this data, Thm.~\ref{ThmUpperBound} now says that there are at most:
\begin{equation}
n_{\alpha,\beta} = \frac{8 \cdot 2 \cdot \frac{7}{4}}{2 \cdot 1} = 14
\end{equation}
interactions.

\begin{figure}
\begin{center}
\epsfig{file=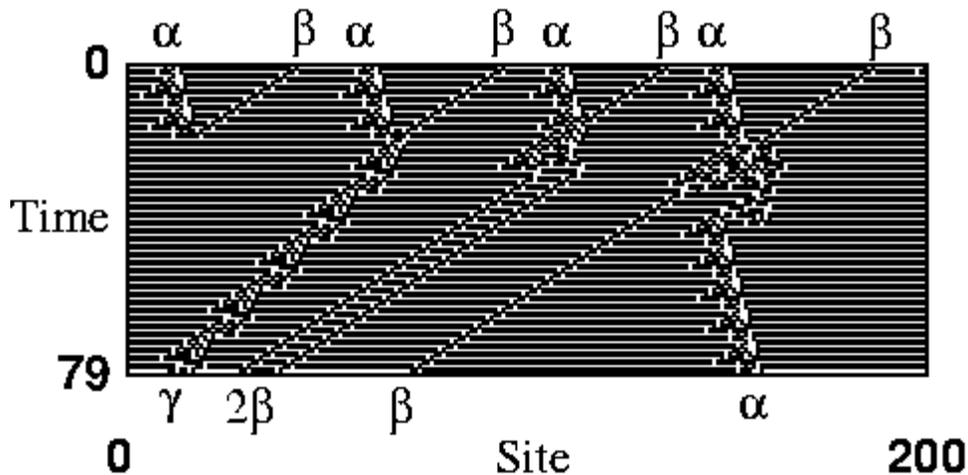,width=5.0in}
\end{center}
\caption{The four different (out of 14 possible) interaction products
  for the $\alpha+\beta$ interaction.}
\label{figure:example2}
\end{figure}

Of these 14 input configurations, it turns out several give rise to the same
products. From a complete survey of $\alpha$-$\Lambda$-$\beta$ configurations,
the result is that there are actually only $4$ different products from the
$\alpha+\beta$ interaction; these are:
\begin{eqnarray*}
\alpha + \beta & \rightarrow & \emptyset \\
\alpha + \beta & \rightarrow & \gamma \\
\alpha + \beta & \rightarrow & 2\beta \\
\alpha + \beta & \rightarrow & \beta +\alpha \\
\end{eqnarray*}
They are shown in Fig.~\ref{figure:example2}.

This example serves to highlight the distinction between the maximum number
of interaction configurations, as bounded by Thm.~\ref{ThmUpperBound},
and the actual number of unique products of the interaction. We shall return
to this distinction later on.

\end{multicols}

\begin{multicols}{2}

\subsection{ECA 110}

In the next example, we test Thm.~\ref{ThmUpperBound} on one of the
long-appreciated ``complex'' CA, elementary CA 110.  As long ago as 1986,
Wolfram \cite[Appendix 15]{Wolfram-theory-and-applications} conjectured that
this rule is able to support universal, Turing-equivalent computation
(replacing an earlier dictum
\cite[p.~31]{Wolfram-universality-and-complexity} that all elementary CA are
``too simple to support universal computation'').  While this conjecture
initially excited little interest, in the last few years it has won increasing
acceptance in the CA research community.  Though to date there is no published
proof of universality, there are studies of its unusually rich variety of
domains and particles, one of the most noteworthy of which is McIntosh's work
on their tiling and tessellation properties \cite{McIntosh-on-110}.  Because of
this CA's behavioral richness, we do not present its complete particle catalog
and computational-mechanical analysis here; rather see
\cite{JPC-CRS-intrinsic-comp}. Instead, we confine ourselves to a single type
of reaction where the utility of our upper bound theorem is particularly
notable.

We consider one domain, labeled $\Lambda^0$, and two particles that move
through it, called $\beta$ and $\kappa$ \cite{JPC-CRS-intrinsic-comp}. (This
$\beta$ particle is not to be confused with the $\beta$ of our previous
examples.)  $\Lambda^0$ is ECA 110's ``true vacuum'': the domain that is stable
and overwhelmingly the most prominent in space-time diagrams generated from
random samples of initial configurations. It has a temporal period
$T(\Lambda^0) = 1$, but a spatial period $S(\Lambda^0) = 14$. The $\beta$
particle has a period $P(\beta) = 15$, during the course of which it moves four
steps to the left: $d_\beta = 4$. The $\kappa$ particle, finally, has a period
$P(\kappa) = 42$, and moves $d_\kappa = 14$ steps to the left during its
cycle. This data gives the $\beta$ particle a velocity of $v_\beta = -4/15$ and
the $\kappa$ particle $v_\kappa = -1/3$.

\begin{figure}
\begin{center}
\epsfig{file=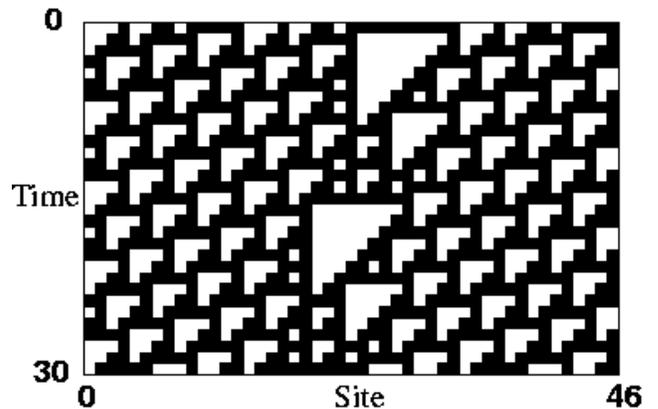,width=3.3in}
\end{center}
\caption{The particle $\beta$ of ECA 110: The space-time patch shows two
  complete cycles of particle phase.}
\label{figure:BetaUnfiltered}
\end{figure}

\begin{figure}
\begin{center}
\epsfig{figure=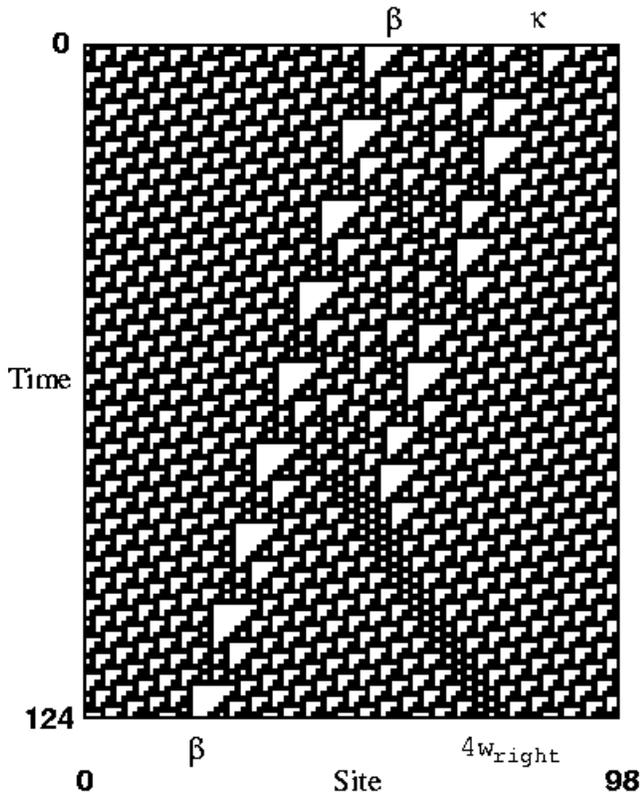,width=3.3in}
\end{center}
\caption{The reaction $\beta + \kappa \rightarrow \beta + 4w_{right}$ in
  ECA 110.}
\label{figure:example4}
\end{figure}

\begin{figure}
\begin{center}
\epsfig{figure=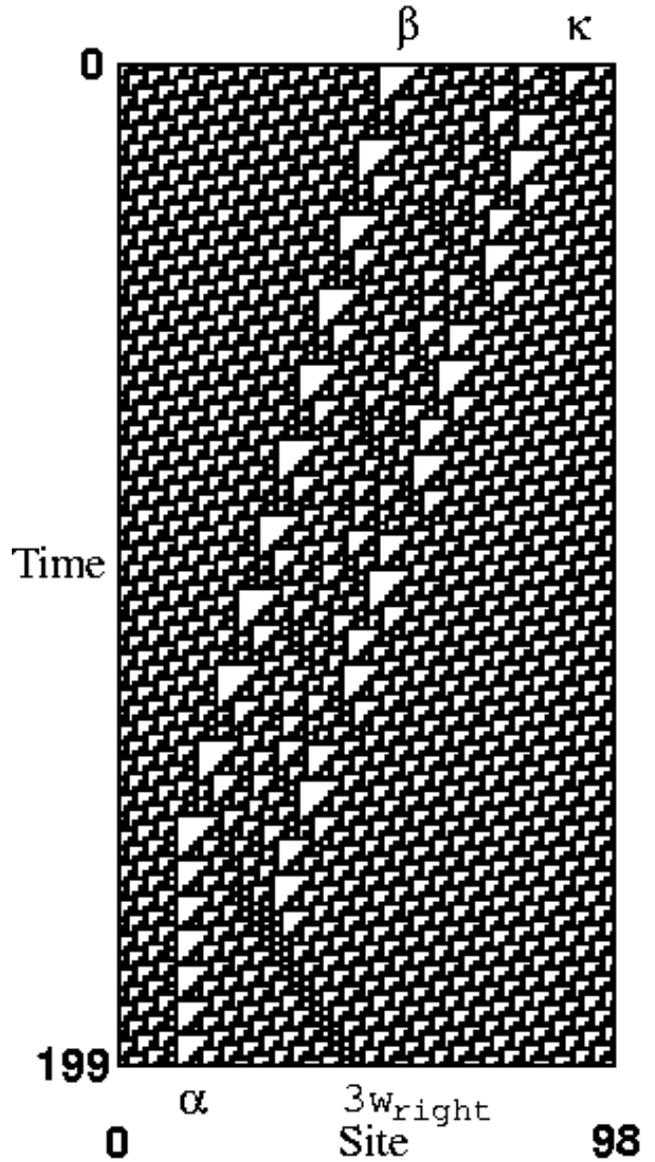,width=3.3in}
\end{center}
\caption{The reaction $\beta + \kappa \rightarrow \alpha + 3w_{right}$ in
  ECA 110.}
\label{figure:example3}
\end{figure}

\begin{figure}[t]
\begin{center}
\epsfig{figure=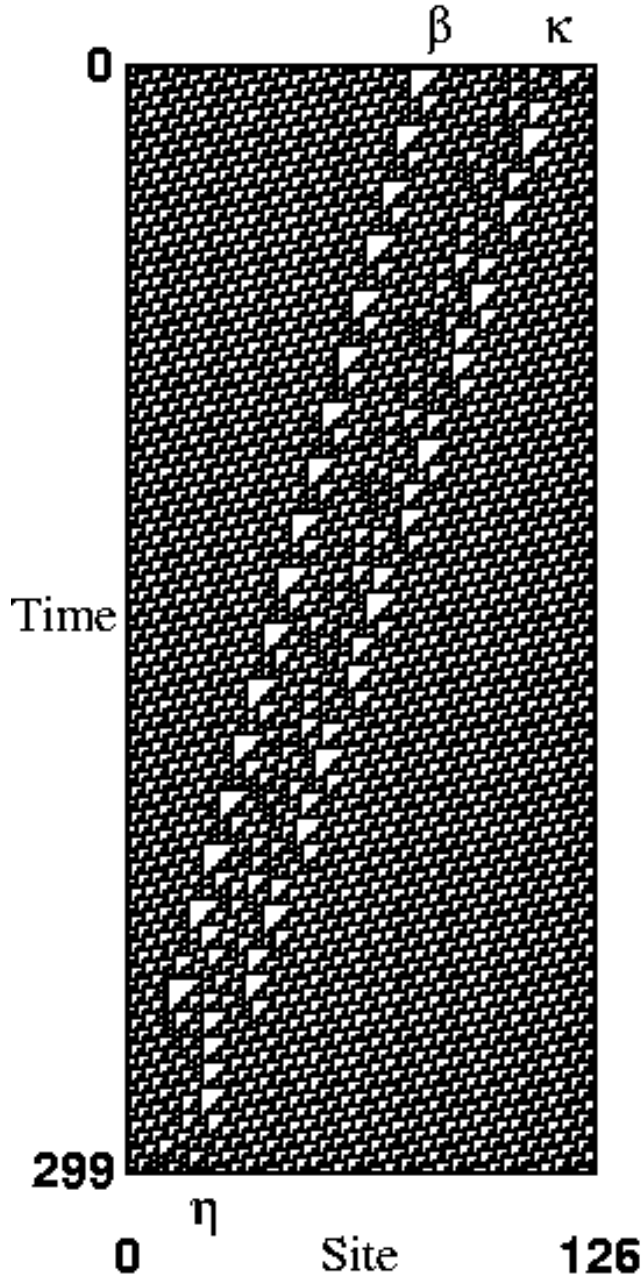,width=3.3in}
\end{center}
\caption{The reaction $ \beta + \kappa \rightarrow \eta$ in ECA 110.}
\label{figure:example5}
\end{figure}

Naively, one would expect to have to examine $630$
($= P(\beta) P(\kappa) = 15 \times 42$) different particle-phase configurations
to exhaust all possible interactions. Theorem \ref{ThmUpperBound}, however,
tells us that all but
\begin{equation}
{(15)(42)({-4 \over 15} - {-1 \over 3}) \over (14)(1)} = 3
\end{equation}
of those initial configurations are redundant. In fact, an exhaustive search
shows that there are exactly three distinct interactions:
\begin{eqnarray*}
\beta + \kappa & \rightarrow & \alpha + 3w_{right} ~, \\
\beta + \kappa & \rightarrow & \beta + 4 w_{right} ~, \\
\beta + \kappa & \rightarrow & \eta ~.
\end{eqnarray*}
Here, $\alpha$, $w_{right}$, and $\eta$ are additional particles generated by
ECA 110. These interactions are depicted, respectively, in Figures
\ref{figure:example3}, \ref{figure:example4}, and \ref{figure:example5}.

We should note that the $w_{right}$ particle is somewhat unusual in that
several can propagate side by side, or even constitute a domain of their own.
There are a number of such ``extensible'' particle families in ECA 110
\cite{JPC-CRS-intrinsic-comp}.

\begin{figure}
\begin{center}
\epsfig{file=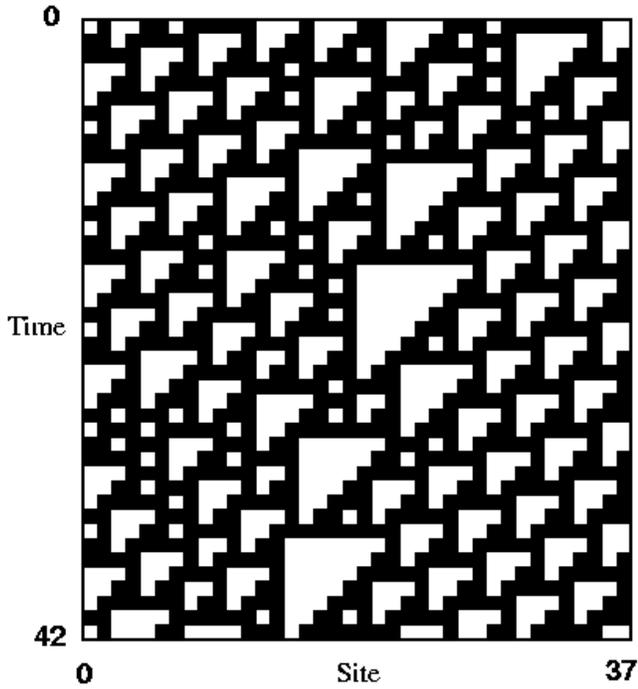,width=3.3in}
\end{center}
\caption{The particle $\kappa$ of ECA 110: The space-time diagram shows
  one complete cycle of particle phase.}
\label{figure:KappaUnfiltered}
\end{figure}

We again see, in this complex case, that the bound of Thm.~\ref{ThmUpperBound}
is attained.

\section{Conclusion}

\subsection{Summary}

The original interaction product formula of
\cite{Park-Steiglitz-Thurston-soliton} is limited to particles propagating in a
completely uniform background; i.e., to a domain whose spatial and temporal
periods are both $1$. When compared to the rich diversity of domains generated
by CAs, this is a considerable restriction, and so the formula does not help in
analyzing many CAs. We have generalized the original result and along the way
established a number of properties of domains and particles---structures
defined by CA computational mechanics.  The examples showed that the upper
bound is tight and that, in complex CAs, particle interactions are
substantially less complicated than they look at first blush. Moreover, in
developing the bound for complex domains, the analysis elucidated the somewhat
subtle notion of a particle's intrinsic periodicity---a property not apparent
from the CA's raw space-time behavior: it requires rather an explicit
representation of the bordering domains' structure.

Understanding the detailed structure of particles and their interactions moves
us closer to an engineering discipline that would tell one how to design CA to
perform a wide range of spatial computations using various particle types,
interactions, and geometries. In a complementary way, it also brings us closer
to scientific methods for analyzing the intrinsic computation of spatially
extended systems \cite{JPC-CRS-intrinsic-comp}.

\subsection{Open Problems}

The foregoing analysis merely scratches the surface of a detailed analytical
approach to CA particle ``physics'': Each CA update rule specifies a
microphysics of local (cell-to-cell) space and time interactions for its
universe; the goal is to discover and analyze those emergent structures that
control the macroscopic behavior.  For now, we can only list a few of the open,
but seemingly accessible, questions our results suggest.

It would be preferable to directly calculate the number of products coming out
of the interaction region, rather than (as here) the number of distinct
particle-domain-particle configurations coming into the interaction region.  We
believe this is eminently achievable, given the detailed representations of
domain and particles that are entailed by a computational mechanics analysis of
CAs.

Two very desirable extensions of these results suggest themselves. The first is
to go from strictly periodic domains to cyclic (periodic and ``chaotic'')
domains and then to general domains. The principle difficulty here is that
Prop.~\ref{PROPSPATIALPERIODICITY} plays a crucial role in our current proof,
but we do not yet see how to generalize its proof to chaotic (positive entropy
density) domains. The second extension would be to incorporate aperiodic
particles, such as the simple one exhibited by ECA 18
\cite{Attractor-Vicinity-Decay}.  We suspect this will prove considerably more
difficult than the extension to cyclic domains: it is not obvious how to apply
notions like ``particle period'' and ``velocity'' to these defects.  A third
extension, perhaps more tractable than the last, is to interactions of more
than two particles.  The geometry and combinatorics will be more complicated
than in the two-particle case, but we conjecture that it will be possible to
establish an upper bound on the number of interaction products for $n-$particle
interactions via induction.

Does there exist an analogous lower bound on the number of interactions?  If
so, when do the upper and lower bounds coincide?

In solitonic interactions the particle number is preserved
\cite{Peyrard-Kruskal,Aizawa-Nishikawa-Kaneko-soliton-turbulence,Park-Steiglitz-Thurston-soliton,Steiglitz-Kamal-Watson,Ablowitz-Kruskal-Ladik}. What
are the conditions on the interaction structure that characterize solitonic
interactions?  The class of soliton-like particles studied in
\cite{Park-Steiglitz-Thurston-soliton} possess a rich ``thermodynamics''
closely analogous to ordinary thermodynamics, explored in detailed in
\cite{Goldberg-parity-filter-automata}.  Do these results generalize to the
broader class of domains and particles, as the original upper bound of
\cite{Park-Steiglitz-Thurston-soliton} does?

While the particle catalog for ECA 110 is not yet {\it provably} complete, for
every known pair of particles the number of distinct interaction products is
exactly equal to the upper bound given by our theorem.  This is not generally
true of most of the CAs we have analyzed and is especially suggestive in light
of the widely-accepted conjecture that the rule is computation universal.  We
suspect that ECA 110's fullness or behavioral flexibility is connected to its
computational power. (Cf. Remark 2 to Thm.~\ref{ThmUpperBound}.)  However, we
have yet to examine other, computation universal CA to see whether they, too,
saturate the bound of our theorem.  One approach to this question would be to
characterize the computational power of systems employing different kinds of
interactions, as is done in \cite{Jakubowski-Steiglitz-Squier} for computers
built from interacting (continuum) solitary waves.

\section*{Acknowledgments}
This work was partially supported under the SFI Computation, Dynamics, and
Learning Program by AFOSR via NSF grant PHY-9970158 and by DARPA under
contract F30602-00-2-0583.

\begin{appendix}

\section*{Proof of Lemma \ref{PROPSPATIALPERIODICITY}}
%\section{Proof of Lemma \ref{PropSpatialPeriodicity}}

\setcounter{lemma}{0}
% Without the set counter, what comes next gets called Proposition 2, even
% though it's identical to Proposition 1!

\begin{lemma}
If a domain $\Lambda^i$ has a periodic phase, then the domain is periodic, and
the spatial periodicities $S(\Lambda^i_j)$ of all its phases $\Lambda^i_j, j =
0, \ldots, p-1,$ are equal.
\end{lemma}

{\it Proof.} The proof consists of two parts. First, and most importantly, it
is proved that the spatial periodicities of the temporal phases of a periodic
domain $\Lambda^i$ cannot increase and that the periodicity of one
phase implies the periodicity of all its successors. Then it follows
straightforwardly that the spatial periodicities have to be equal for all
temporal phases and that they all must be periodic.

Our proof employs the update transducer $T_{\phi}$, which is simply the FST
which scans across a lattice configuration and outputs the effect of applying
the CA update rule $\phi$ to it.  For reasons of space, we refrain from giving
full details on this operator---see rather \cite{Hanson-thesis}.  Here we need
the following results.  If $\phi$ is a binary, radius-$r$ CA, the update
transducer has $2^{2r}$ states, representing the $2^{2r}$ distinct contexts
(words of previously read symbols) in which $T_{\phi}$ scans new sites, and we
customarily label the states by these context words. The effect of applying the
CA $\phi$ to a set of lattice configuration represented by the DFA $M$ is a new
machine, given by $T_{\phi}M$---the ``direct product'' of the machines $M$ and
$T_{\phi}$.  Once again, for reasons of space, we will not explain how this
direct product works in the general case.  We are interested merely in the
special case where $M = \Lambda^i_j$, the $j^{th}$, periodic phase of a domain,
with spatial period $n$. The next phase of the domain, $\Lambda^i_{j+1}$, is
the composed automaton $T_{\phi}M$, \textit{once the latter has been
minimized}. Before the latter step $T_{\phi}M$ consists of $n$ ``copies'' of
the FST $T_{\phi}$, one for each of $\Lambda^i_j$'s $n$ states. There are no
transitions within a copy.  Transitions from copy $k$ to copy $k^\prime$ occur
only if $k^\prime = k + 1$ ($\bmod ~ n$). In total, there are $n{2}^{2r}$
states in the direct composition.

$T_{\phi}M$ is finite and deterministic, but far from minimal. We are
interested in its minimal equivalent machine, since that is what we have
defined as the representative of the next phase of the domain.  The key to our
proof is an unproblematic part of the minimization, namely, removing states
that have no predecessors (i.e., no incoming transitions) and so are never
reached. (Recall that, by hypothesis, we are examining successive phases of a
domain, all represented by strongly connected process graphs.) It can be shown,
using the techniques in \cite{Hanson-thesis}, that if the transition from state
$k$ in $\Lambda^i_j$ to state $k+1$ occurs on a $0$ (respectively, on a $1$),
then in the composed machine, the transitions from copy $k$ of $T_{\phi}$ only
go to those states in copy $k+1$ whose context string ends in a $0$
(respectively, in a $1$). Since states in copy $k+1$ can be reached only from
states in copy $k$, it follows that half of the states in each copy cannot be
reached at all, and so they can be eliminated without loss.

Now, this procedure of eliminating states without direct predecessors in turn
leaves some states in copy $k+2$ without predecessors. So we can re-apply the
procedure, and once again, it will remove half of the remaining states.  This
is because applying it twice is the same as removing those states in copy $k+2$
for which the last two symbols in the context word differ from the symbols
connecting state $k$ to state $k+1$ and state $k+1$ to state $k+2$ in the
original domain machine $\Lambda^i_j$.

What this procedure does is exploit the fact that, in a domain, every state is
encountered only in a unique update-scanning context; we are eliminating
combinations of domain-state and update-transducer-state that simply cannot be
reached.  Observe that we can apply this procedure exactly $2r$ times, since
that suffices to establish the complete scanning context, and each time we do
so, we eliminate half the remaining states. We are left then with
$n{2}^{2r}/2^{2r} = n$ states after this process of successive halvings.
Further observe that, since each state $k$ of the original domain machine
$\Lambda^i_j$ occurs in \textit{some} scanning context, we will never eliminate
\textit{all} the states in copy $k$.  Since each of the $n$ copies has at least
one state left in it, and there are only $n$ states remaining after the
halvings are done, it follows that each copy contains exactly one state, which
has one incoming transition, from the previous copy, and one outgoing
transition, to the next copy.  The result of eliminating unreachable states,
therefore, is a machine of $n$ states which is not just deterministic but (as
we have defined the term) periodic.  Note, however, that this is not
\textit{necessarily} the minimal machine, since we have not gone through a
complete minimization procedure, merely the easy part of one.
$\Lambda^i_{j+1}$ thus might have fewer than $n$ states, but certainly no more.

To sum up: We have established that, if $\Lambda^i_j$ is a periodic domain
phase, then $\Lambda^i_{j+1}$ is also periodic and
$S(\Lambda^i_{j+1}) \leq S(\Lambda^i_j)$. Thus, for any $t$,
$S( \boldsymbol{\Phi}^t \Lambda^i_j) \leq S(\Lambda^i_j)$. But
$\boldsymbol{\Phi}^t \Lambda^i_j) = \Lambda^i_{(j+t) \bmod p}$
and if $t = p$, we have
$\Lambda^i_{(j+t) \bmod p} = \Lambda^i_{(j+p)\bmod p} = \Lambda^i_j$.
Putting these together we have
\begin{equation}
S(\Lambda^i_{j+1}) \leq S(\Lambda^i_j)
  \Rightarrow S(\Lambda^i_{j+1}) = S(\Lambda^i_j) ~,
\end{equation}
for $j = 0, 1, \ldots, p-1$.  This implies that the spatial period is the same,
namely $n$, for all phases of the domain. And this proves the proposition when
the CA alphabet is binary.

The reader may easily check that a completely parallel argument holds if the CA
alphabet is not binary but $m-$ary, substituting $m$ for 2 and $(m-1)/m$ for
$1/2$ in the appropriate places.  We omit it here for reasons of space and
notational complexity. \hfill$\Box$

\end{appendix}

\bibliography{locusts}

\begin{thebibliography}{10}
\expandafter\ifx\csname url\endcsname\relax
  \def\url#1{\texttt{#1}}\fi
\expandafter\ifx\csname urlprefix\endcsname\relax\def\urlprefix{URL }\fi

\bibitem{Burks-essays}
A.~W. Burks (Ed.), Essays on Cellular Automata, University of Illinois Press,
  Urbana, 1970.

\bibitem{Winning-Ways}
E.~R. Berlekamp, J.~H. Conway, R.~K. Guy, Winning Ways for your Mathematical
  Plays, Academic Press, New York, 1982.

\bibitem{Peyrard-Kruskal}
M.~Peyrard, M.~D. Kruskal, Kink dynamics in the highly discrete sine-{Gordon}
  system, Physica D 14 (1984) 88--102.

\bibitem{Grassberger-diffusion}
P.~Grassberger, New mechanism for deterministic diffusion, Physical Review A 28
  (1983) 3666--7.

\bibitem{Boccara-Nasser-Roger-particle-like}
N.~Boccara, J.~Nasser, M.~Roger, Particle-like structures and their
  interactions in spatio-temporal patterns generated by one-dimensional
  deterministic cellular automaton rules, Physical Review A 44 (1991) 866 --
  875.

\bibitem{Boccara-Roger-block-transformations}
N.~Boccara, M.~Roger, Block transformations of one-dimensional deterministic
  cellular automata, Journal of Physics A 24 (1991) 1849 -- 1865.

\bibitem{Boccara-transformations}
N.~Boccara, Transformations of one-dimensional cellular automaton rules by
  translation-invariant local surjective mappings, Physica D 68 (1993)
  416--426.

\bibitem{Aizawa-Nishikawa-Kaneko-soliton-turbulence}
Y.~Aizawa, I.~Nishikawa, K.~Kaneko, Soliton turbulence in cellular automata,
  in: H.~Gutowitz (Ed.), Cellular Automata: {Theory} and Experiment, MIT Press,
  Cambridge, Massachusetts, 1991, pp. 307--327, also published as
  \textit{Physica D} \textbf{45} (1990), nos. 1--3.

\bibitem{Park-Steiglitz-Thurston-soliton}
J.~K. Park, K.~Steiglitz, W.~P. Thurston, Soliton-like behavior in automata,
  Physica D 19 (1986) 423--432.

\bibitem{Wolfram-theory-and-applications}
S.~Wolfram (Ed.), Theory and Applications of Cellular Automata, World
  Scientific, Singapore, 1986.

\bibitem{Wolfram-CA-and-complexity}
S.~Wolfram, Cellular Automata and Complexity: {Collected} Papers,
  Addison-Wesley, Reading, Massachusetts, 1994, online at
  \texttt{http://www.stephenwolfram.com/publications/books/ ca-reprint/}.

\bibitem{Lindgren-Nordahl-universal-computation}
K.~Lindgren, M.~G. Nordahl, Universal computation in a simple one-dimensional
  cellular automaton, Complex Systems 4 (1990) 299--318.

\bibitem{JPC-MM-PNAS}
J.~P. Crutchfield, M.~Mitchell, The evolution of emergent computation,
  Proceedings of the National Academy of Sciences 92 (1995) 10742--10746.

\bibitem{Yunes-firing-squad}
J.~B. Yunes, Seven-state solutions to the firing squad synchronization problem,
  Theoretical Computer Science 127 (1994) 313--332.

\bibitem{Eloranta-partially-permutive}
K.~Eloranta, Partially permutive cellular automata, Nonlinearity 6 (1993)
  1009--1023.

\bibitem{Eloranta-defect-ensembles}
K.~Eloranta, The dynamics of defect ensembles in one-dimensional cellular
  automata, Journal of Statistical Physics 76 (1994) 1377--1398.

\bibitem{Eloranta-Nummelin-random-walk}
K.~Eloranta, E.~Nummelin, The kink of cellular automaton rule 18 performs a
  random walk, Journal of Statistical Physics 69 (1992) 1131--1136.

\bibitem{Cellular-Automata-and-Modeling}
P.~Manneville, N.~Boccara, G.~Y. Vichniac, R.~Bidaux (Eds.), Cellular Automata
  and Modeling of Complex Systems: {Proceedings} of the Winter School, Les
  Houches, France, February 21--28, 1989, Vol.~46 of Springer Proceedings in
  Physics, Springer-Verlag, Berlin, 1990.

\bibitem{Andre-Bennett-Koza}
D.~Andre, F.~H. Bennett, {III}, J.~R. Koza, Evolution of intricate
  long-distance communication signals in cellular automata using genetic
  programming, in: C.~G. Langton, K.~Shimohara (Eds.), Artificial Life {V}, MIT
  Press, Cambridge, Massachusetts, 1997, pp. 513--520.

\bibitem{Attractor-basin-portrait}
J.~E. Hanson, J.~P. Crutchfield, The attractor-basin portrait of a cellular
  automaton, Journal of Statistical Phyics 66 (1992) 1415--1462.

\bibitem{Hanson-thesis}
J.~E. Hanson, Computational mechanics of cellular automata, Ph.D. thesis,
  University of California, Berkeley (1993).

\bibitem{Comp-mech-of-CA-example}
J.~E. Hanson, J.~P. Crutchfield, Computational mechanics of cellular automata:
  {A}n example, Physica D 103 (1997) 169--189.

\bibitem{Eppstein-glider-rule-database}
D.~Eppstein, Gliders in life-like cellular automata, Interactive online
  database of two-dimensional cellular automata rules with gliders,
  \texttt{http://fano.ics.uci.edu/ca/}.

\bibitem{Manneville-dissipative-structures}
P.~Manneville, Dissipative Structures and Weak Turbulence, Academic Press,
  Boston, Massachusetts, 1990.

\bibitem{Cross-Hohenberg}
M.~C. Cross, P.~Hohenberg, Pattern {F}ormation {O}ut of {E}quilibrium, Reviews
  of Modern Physics 65 (1993) 851--1112.

\bibitem{Winfree-geometry}
A.~T. Winfree, The Geometry of Biological Time, Springer-Verlag, Berlin, 1980.

\bibitem{Winfree-time-breaks-down}
A.~T. Winfree, When Time Breaks Down: {T}he Three-Dimensional Dynamics of
  Electrochemical Waves and Cardiac Arrhythmias, Princeton University Press,
  Princeton, 1987.

\bibitem{Infeld-Rowlands}
E.~Infeld, G.~Rowlands, Nonlinear Waves, Solitions, and Chaos, Cambridge
  University Press, Cambridge, England, 1990.

\bibitem{Poundstone-recursive}
W.~Poundstone, The Recursive Universe: {Cosmic} Complexity and the Limits of
  Scientific Knowledge, William Morrow, New York, 1984.

\bibitem{Griffeath-particle-systems}
D.~Griffeath, Additive and Cancellative Interacting Particle Systems, Vol. 724
  of Lecture Notes in Mathematics, Springer-Verlag, Berlin, 1979.

\bibitem{Liggett-particle-systems}
T.~M. Liggett, Interacting Particle Systems, Springer-Verlag, Berlin, 1985.

\bibitem{Rothman-Zaleski-text}
D.~H. Rothman, S.~Zaleski, Lattice-Gas Cellular Automata: Simple Models of
  Complex Hydrodynamics, Vol.~5 of Collection Al{\'e}a Saclay, Cambridge
  University Press, Cambridge, England, 1997.

\bibitem{Steiglitz-Kamal-Watson}
K.~Steiglitz, I.~Kamal, A.~Watson, Embedding computation in one-dimensional
  automata by phase coding solitons, IEEE Transactions on Computers 37 (1988)
  138--144.

\bibitem{Griffeath-Moore-LwoD}
D.~Griffeath, C.~Moore, Life without death is {P}-complete, Complex Systems 10
  (1996) 437--447.

\bibitem{Moore-majority-vote}
C.~Moore, Majority-vote cellular automata, {Ising} dynamics, and
  {P}-completeness, Journal of Statistical Physics 88 (1997) 795--805.

\bibitem{Moore-Nordahl-lattice-gases}
C.~Moore, M.~G. Nordahl, Lattice gas prediction is {P}-complete, Electronic
  pre-preprint, arxiv.org, comp-gas/9704001 (1997).

\bibitem{Das-MM-JPC-discovery-of-particles}
R.~Das, M.~Mitchell, J.~P. Crutchfield, A genetic algorithm discovers particle
  computation in cellular automata, in: Y.~Davidor, H.-P. Schwefel, R.~Manner
  (Eds.), Proceedings of the Conference on Parallel Problem Solving in Nature
  --- PPSN III, Lecture Notes in Computer Science, Springer-Verlag, Berlin,
  1994, pp. 344--353.

\bibitem{Wim-MM-JPC-mechanisms}
W.~Hordijk, M.~Mitchell, J.~P. Crutchfield, Mechanisms of emergent computation
  in cellular automata, in: A.~E. Eiben, T.~B{\"a}ck, M.~Schoenaur, H.-P.
  Schwefel (Eds.), Parallel Problem Solving in Nature---PPSN V, Lecture Notes
  in Computer Science, Springer-Verlag, Berlin, 1998, pp. 613--622.

\bibitem{Margolus-crystalline}
N.~Margolus, Crystalline computation, in: A.~J.~G. Hey (Ed.), Feynman and
  Computation: {Exploring} the Limits of Computers, Perseus Books, Reading,
  Massachusetts, 1999, pp. 267--305, e-print, arxiv.org, comp-gas/9811002.

\bibitem{Das-thesis}
R.~Das, The evolution of emergent computation in cellular automata, Ph.D.
  thesis, Colorado State University (1996).

\bibitem{Wim-thesis}
W.~Hordijk, Dynamics, emergent computation, and evolution in cellular automata,
  Ph.D. thesis, University of New Mexico, Albuquerque, New Mexico, online at
  \texttt{http://www.santafe.edu/projects/ evca/ Papers/ WH-Diss.html} (1999).

\bibitem{Wolfram-universality-and-complexity}
S.~Wolfram, Universality and complexity in cellular automata, Physica D 10
  (1984) 1--35, reprinted in \cite{Wolfram-CA-and-complexity}.

\bibitem{Anything-ever-new}
J.~P. Crutchfield, Is anything ever new? {C}onsidering emergence, in: G.~Cowan,
  D.~Pines, D.~Melzner (Eds.), Complexity: {M}etaphors, Models, and Reality,
  Vol.~19 of Santa Fe Institute Studies in the Sciences of Complexity,
  Addison-Wesley, Reading, Massachusetts, 1994, pp. 479--497.

\bibitem{Holland-emergence}
J.~H. Holland, Emergence: {F}rom Chaos to Order, Addison-Wesley, Reading,
  Massachusetts, 1998.

\bibitem{Wuesnche-glider-finding}
A.~Wuensche, Classifying cellular automata automatically: {Finding} gliders,
  filtering, and relating space-time patterns, attractor basins, and the {Z}
  parameter, Complexity 4 (1999) 47--66.

\bibitem{Eppstein-searching-for-spaceships}
D.~Eppstein, Searching for spaceships, E-print, arxiv.org, cs.AI/0004003
  (2000).

\bibitem{Wolfram-stat-mech-CA}
S.~Wolfram, Statistical mechanics of cellular automata, Reviews of Modern
  Physics 55 (1983) 601--644, reprinted in \cite{Wolfram-CA-and-complexity}.

\bibitem{Wolfram-computation}
S.~Wolfram, Computation theory of cellular automata, Communications in
  Mathematical Physics 96 (1984) 15--57, reprinted in
  \cite{Wolfram-CA-and-complexity}.

\bibitem{Hopcroft-Ullman}
J.~E. Hopcroft, J.~D. Ullman, Introduction to Automata Theory, Languages, and
  Computation, Addison-Wesley, Reading, 1979, 2nd edition of {\it Formal
  Languages and Their Relation to Automata}, 1969.

\bibitem{Inferring-stat-compl}
J.~P. Crutchfield, K.~Young, Inferring statistical complexity, Physical Review
  Letters 63 (1989) 105--108.

\bibitem{Turbulent-pattern-bases}
J.~P. Crutchfield, J.~E. Hanson, Turbulent pattern bases for cellular automata,
  Physica D 69 (1993) 279--301.

\bibitem{Burton-number-theory}
D.~M. Burton, Elementary Number Theory, Allyn and Bacon, Boston, 1976.

\bibitem{JPC-Wim-MM-performance}
J.~P. Crutchfield, W.~Hordijk, M.~Mitchell, Computational performance of
  evolved cellular automata: {Parts} {I} and {II}, manuscript in preparation.

\bibitem{McIntosh-on-110}
H.~V. McIntosh, Rule 110 as it relates to the presence of gliders, Electronic
  manuscript, \texttt{http://delta.cs.cinvestav.mx/$\sim$mcintosh/comun/
  RULE110W/RULE110.html} (2000).

\bibitem{JPC-CRS-intrinsic-comp}
J.~P. Crutchfield, C.~R. Shalizi, Intrinsic computation versus engineered
  computation: {The} computational mechanics of rule 110, manuscript in
  preparation.

\bibitem{Attractor-Vicinity-Decay}
J.~P. Crutchfield, J.~E. Hanson, Attractor vicinity decay for a cellular
  automaton, Chaos 3 (1993) 215--224.

\bibitem{Ablowitz-Kruskal-Ladik}
M.~J. Ablowitz, M.~D. Kruskal, J.~F. Ladik, Solitary wave collisons, SIAM
  Journal on Applied Mathematics 36 (1979) 428--437.

\bibitem{Goldberg-parity-filter-automata}
C.~H. Goldberg, Parity filter automata, Complex Systems 2 (1988) 91--141.

\bibitem{Jakubowski-Steiglitz-Squier}
M.~H. Jakubowski, K.~Steiglitz, R.~Squier, Information transfer between
  solitary waves in the saturable {S}chr{\"o}dinger equation, Physical Review E
  56 (1997) 7267--7272.

\end{thebibliography}
\bibliographystyle{elsart-num}

\end{multicols}

\end{document}